\documentclass[prd,aps, preprintnumbers, showpacs,twocolumn, nofootinbib,superscriptaddress,notitlepage]{revtex4-1}
\usepackage{stmaryrd}
\usepackage[normalem]{ulem}
\usepackage{epsfig}
\usepackage{amsfonts}
\usepackage{amsmath}
\usepackage{slashed}
\usepackage{graphicx}
\usepackage{color}
\usepackage{mathtools}
\usepackage{subfigure}
\usepackage{graphicx}
\usepackage{float}
\usepackage{placeins}

\usepackage{amssymb,psfrag}
\usepackage{caption}
\usepackage{float}
\usepackage[table]{xcolor}
\usepackage{verbatim}

\newcommand{\beq}{\begin{eqnarray}}
	\newcommand{\eeq}{\end{eqnarray}}

\newcommand{\nn}{\nonumber}

\begin{document}
\title{Testing a Linear Relation: Short-Range Correlations and the EMC Effect for Gluons and Quarks in Nuclei}
	
\author{Shu-Man Hu}
\affiliation{Frontiers Science Center for Rare Isotopes, and School of Nuclear Science and Technology, Lanzhou University, Lanzhou 730000, China}
	
\author{Wei Wang}
\email{wei.wang@sjtu.edu.cn}
\affiliation{INPAC, Key Laboratory for Particle Astrophysics and Cosmology (MOE), Shanghai Key Laboratory for Particle Physics and Cosmology,
School of Physics and Astronomy, Shanghai Jiao Tong University, Shanghai 200240, China}

\author{Ji Xu}
\email{xuji@lzu.edu.cn}
\affiliation{Frontiers Science Center for Rare Isotopes, and School of Nuclear Science and Technology, Lanzhou University, Lanzhou 730000, China}
	
\author{Xing-Hua Yang}
\email{yangxinghua@sdut.edu.cn}
\affiliation{School of Physics and Optoelectronic Engineering, Shandong University of Technology, Zibo, Shandong 255000, China}

\author{Shuai Zhao}
\email{zhaos@tju.edu.cn}
\affiliation{Department of Physics, School of Science, Tianjin University, Tianjin 300350, China}

\date{\today}
	
\begin{abstract}
  In this work, we focus on the possible linear relation between short-range correlations (SRCs) and the EMC effect for partons in nuclei. First, we test a linear relationship pertaining to gluons in bound nuclei; it is manifested as a correlation between the slope of the reduced cross section ratio in deep inelastic scattering (DIS) and the cross section of sub-threshold $J/\psi$ photoproduction. For comparison, the results from four different global analyses groups of nuclear parton distribution functions (nPDFs) are utilized. These results show a good linear correlation between the gluons in bound nuclei and the slope of the reduced cross section ratio, consistent with the possible presence of nuclear effects in the gluon distributions. Second, we investigate the linear relationship of quarks in the proton-induced Drell-Yan process. The corresponding results for quarks show strong sensitivity to the parameterization forms adopted by the different groups. These findings enhance our understanding of the substructure in bound nuclei and provide valuable reference for future global fitting of nPDFs.
\end{abstract}
	
	\maketitle
	
\section{Introduction}
The nuclei consist of strongly interacting protons and neutrons, which give rise to the vast majority of mass in the visible universe. Inclusive lepton scattering is an effective tool for studying nuclei. Previously, it was widely believed that the quark structure within nucleons would not be influenced by the structure of atomic nuclei due to energy scale separation. However, the European Muon Collaboration discovered that the per-nucleon deep inelastic scattering (DIS) cross section of iron to deuteron is degressive at $0.3\leq x \leq 0.7$, which is now known as the EMC effect \cite{EuropeanMuon:1983wih}. Since then, the EMC effect has been confirmed in different nuclei \cite{EuropeanMuon:1988lbf,Gomez:1993ri,EuropeanMuon:1988tpw,NewMuonNMC:1990xyw,Seely:2009gt,CLAS:2019vsb}, indicating that the quark structure is significantly modified by the nuclear medium. The origin of EMC effect has numerous explanations \cite{Wang:2022kwg,Bertsch:1993vx,Miller:2013hla,Frank:1995pv,Wang:2016mzo,Zhang:2009vj}, yet a consensus remains elusive. Recent studies considering the relation between the EMC effect and nucleon-nucleon short range correlations (SRCs) have attracted a lot of attentions \cite{Weinstein:2010rt,CLAS:2005ola,Wang:2024ikx,Hen:2012fm,Hen:2014nza,CLAS:2018yvt}. The SRC is strongly interacting nucleon pair with a large relative momentum and a small center-of-mass momentum in comparison to the single-nucleon Fermi momentum, and this short range interaction may cause a modification of the structure of the nucleon.

The authors in Ref.\,\cite{Weinstein:2010rt} quantified the relationship between these two effects through a linear correlation between the SRC scaling factor and the size of the EMC effect. In line with this idea, a similar linear correlation was proposed between the heavy flavor production in DIS and sub-threshold photoproduction of $J/\psi$, which are two gluon-centric progresses \cite{Wang:2024cpx}. It is noteworthy that, while some studies regard SRCs as a key transitional structure connecting quark-gluon and nucleonic degrees of freedoms \cite{nCTEQ:2023cpo}, others contend that current experimental data are also compatible with non-SRC interpretive frameworks \cite{Paakkinen:2025pcw}.

Therefore, a better understanding of the nuclear parton (quark and gluons) distributions and SRCs requires more careful examinations \cite{Xu:2019wso,Hatta:2019ocp,Chen:2016bde,Wang:2020uhj,Yang:2023zmr,Huang:2021cac}. The linear relation presented in Ref.\,\cite{Wang:2024cpx} relies on the global analysis of gluon nuclear parton distribution functions (nPDFs) in EPPS21 \cite{Eskola:2021nhw}. When extracting the nPDFs, different collaboration groups may adopt different experimental data and parameterization forms. Would this linear relationship still hold across different groups? Currently, a feasible approach is to collect and compare all available fitting results for gluon nPDFs in the market. In this work, we present a detailed analysis on this issue by employing four commonly used nPDFs parameterizations: EPPS21\footnote{https://research.hip.fi/qcdtheory/nuclear-pdfs/epps21/} \cite{Eskola:2021nhw}, nNNPDF3.0(no LHCb D)\footnote{https://nnpdf.mi.infn.it/for-users/nnnpdf3-0/} \cite{AbdulKhalek:2022fyi}, nCTEQ15HQ\footnote{https://ncteq.hepforge.org/ncteq15hq/index.html} \cite{Duwentaster:2022kpv} and TUJU21\footnote{https://lhapdf.hepforge.org/pdfsets.html} \cite{Helenius:2021tof}. 
Furthermore, most of these datasets are also incorporated in the LHAPDF library \cite{Buckley:2014ana}.

The study of the EMC effect through proton-induced Drell-Yan process holds unique importance as a complementary probe to DIS. By selecting kinematic regions sensitive to quark distributions of the nuclear target, these experiments can test the specific suppression of quarks in bound nucleons (EMC effect) \cite{NuSea:1999egr,NA3:1981yaj,NA10:1987hho,Reimer:2016dcd,Reimer:2011zza}. In this work, we also investigate the linear relationship of quarks in the proton-induced Drell-Yan process by utilizing the nPDFs of these four groups.

This paper is organized as follows. In Sec.\,\ref{lineargluon}, the linear relation of gluons proposed in DIS and sub-threshold photoproduction of $J/\psi$ processes are tested. In Sec.\,\ref{linearquark}, we examine the linear relation of quarks in proton-induced Drell-Yan process. Sec.\,\ref{summary} is reserved for conclusions. Some additional tables and figures are collected in the appendices.

\section{Test of linear relation of gluons}
\label{lineargluon}
Exploring gluon EMC effects requests the knowledge of the nPDFs pertaining to gluons. The construction of electron-ion collider (EIC) with a possibility to operate with wide variety of nuclei, will constrain the gluon density in nuclei via measurements of the charm reduced cross section in DIS \cite{Aschenauer:2017oxs},
\begin{align}\label{reduceCS}
 &\sigma_{A,red}^{c \bar c}(x, Q^2) \equiv \left(\frac{d\sigma^{c \bar c}_A}{dx dQ^2}\right)\frac{xQ^4}{2\pi \alpha^2[1+(1-y)^2]} \nn\\
 &= \frac{2}{[1+(1-y)^2]}\Big( xy^2 F_{1, A}^{c \bar c}(x,Q^2) + (1-y)F_{2, A}^{c \bar c}(x,Q^2) \Big) \,,\nn\\
\end{align}
here $\sigma^{c \bar c}_A$ denotes the charm cross section which is customarily expressed as the reduced cross section $\sigma_{A,red}^{c \bar c}$; $x$ and $y$ are Bjorken variable and inelasticity, respectively. In the parton model, information about the gluon nPDFs is encoded in the $F_{1/2, A}^{c \bar c}$, which are called charm structure functions,
\begin{eqnarray}\label{final_SF12_integral_expression}
  F_{1,A}^{c \bar c}(x,Q^2) &=& \int_{\tau x}^1 \frac{dz}{z} f_g^A(z,\hat s) \, f_1(\frac{x}{z},Q^2) \,,\nn\\
  F_{2,A}^{c \bar c}(x,Q^2) &=& \int_{\tau x}^1 \frac{dz}{z} z \, f_g^A(z,\hat s) \, f_2(\frac{x}{z},Q^2) \,.
\end{eqnarray}
Here $f_g^A$ is the gluon nPDF in a given nucleus $A$; $f_{1/2}$ represents the perturbative  partonic cross sections whose expressions can be found in \cite{Laenen:1992zk}.

\begin{figure*}[]
	\centering
	\subfigure[EPPS21]{
		\label{RAcc_EPPS21}
		\includegraphics[width=0.44\linewidth, trim=20 20 20 20, clip]{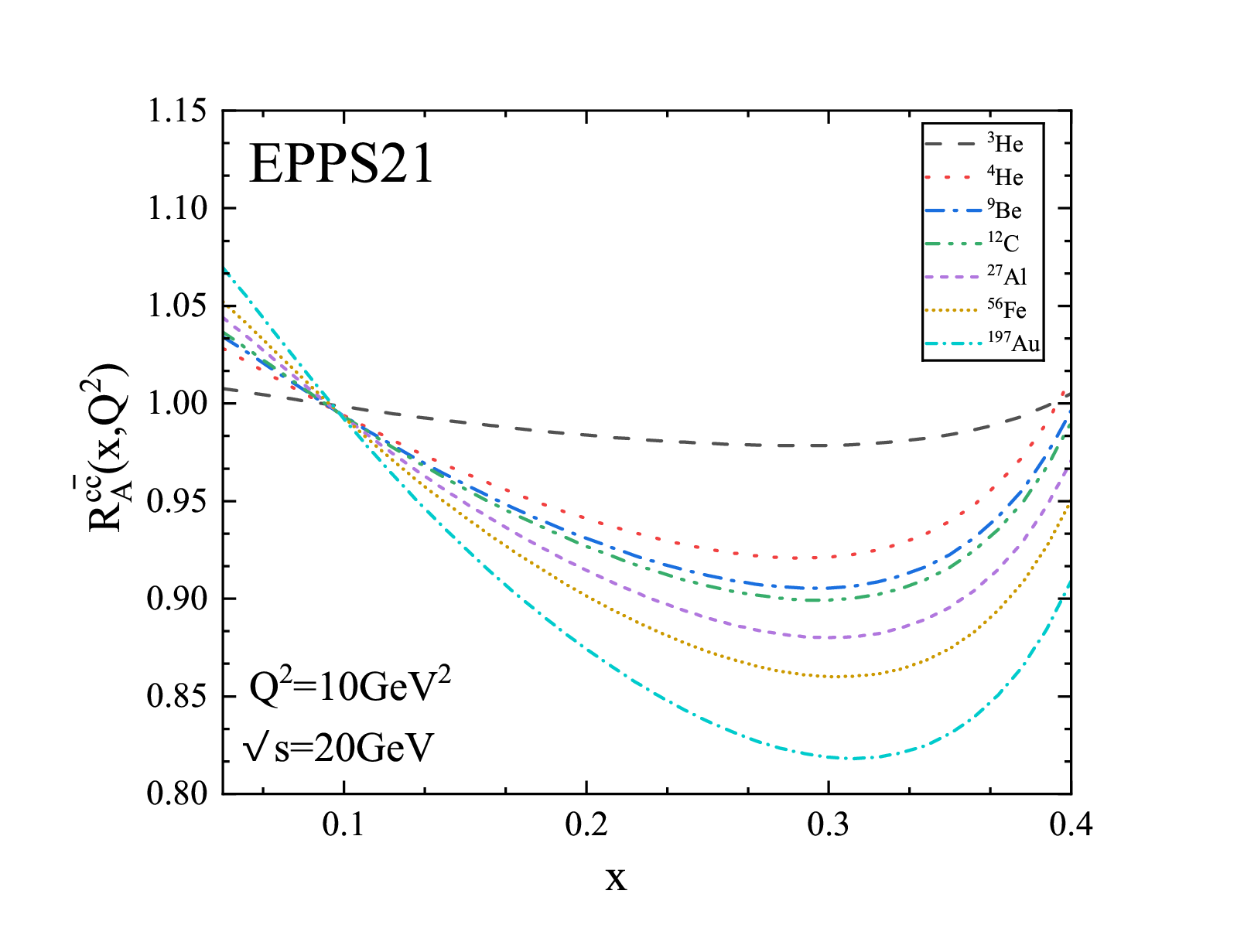}}
	\hspace{-20pt}  
	\subfigure[nNNPDF3.0(no LHCb D)]{
		\label{RAcc_nNNPDF30}
		\includegraphics[width=0.44\linewidth, trim=20 20 20 20, clip]{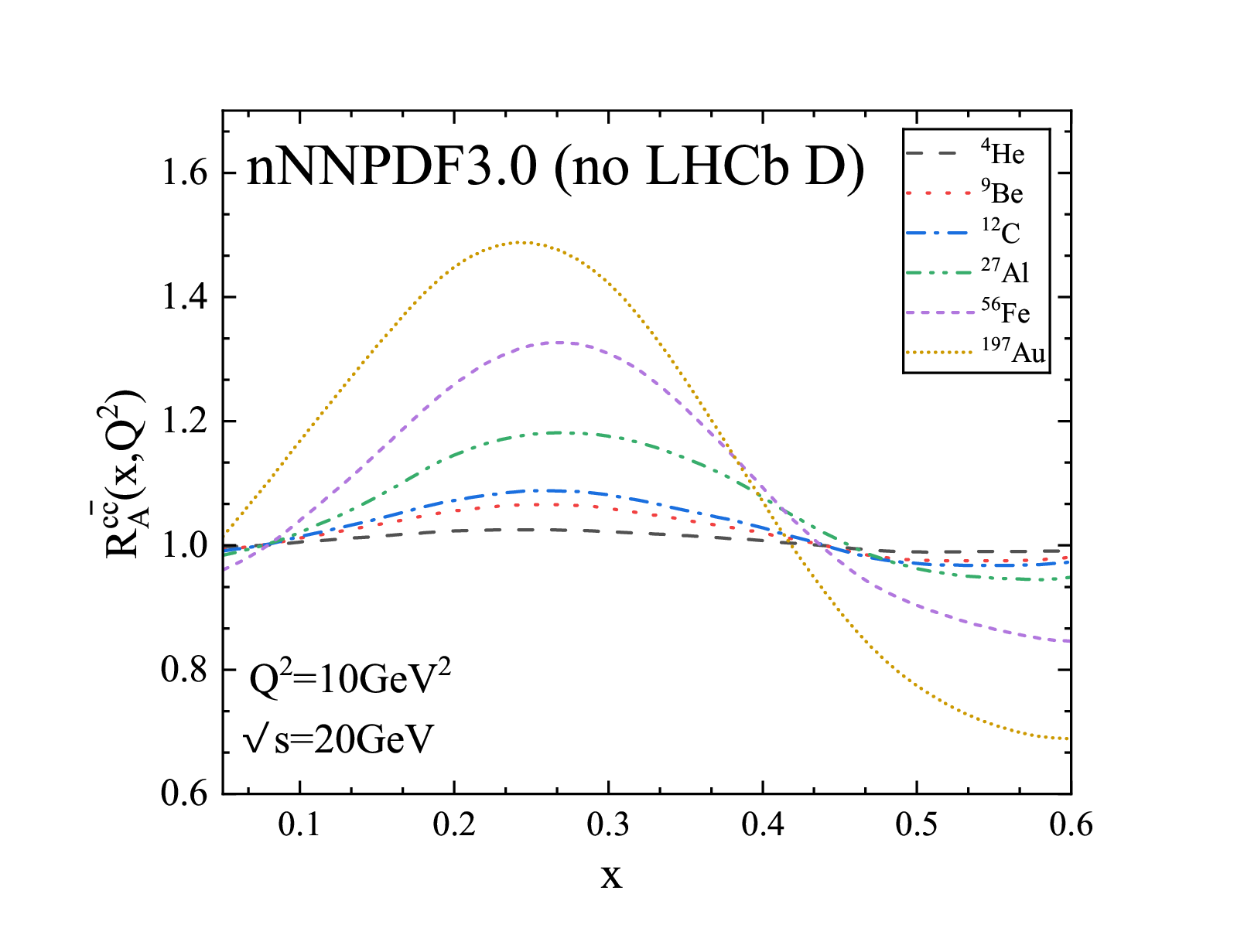}}
	
	\vspace{-8pt}  
	
	\subfigure[TUJU21]{
		\label{RAcc_TUJU21}
		\includegraphics[width=0.44\linewidth, trim=20 20 20 20, clip]{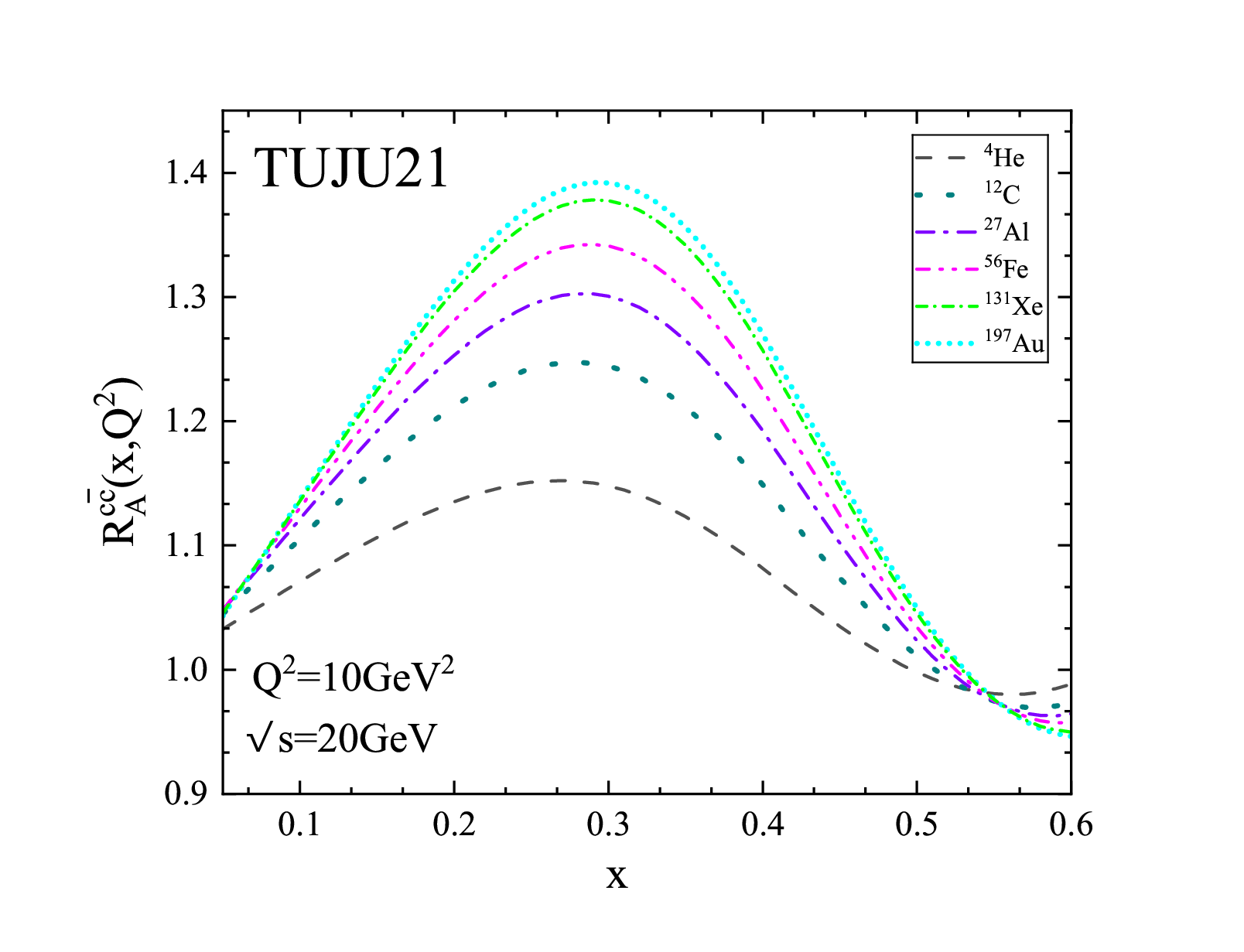}}
	\hspace{-20pt}
	\subfigure[nCTEQ15HQ]{
		\label{RAcc_nCTEQ}
		\includegraphics[width=0.44\linewidth, trim=20 20 20 20, clip]{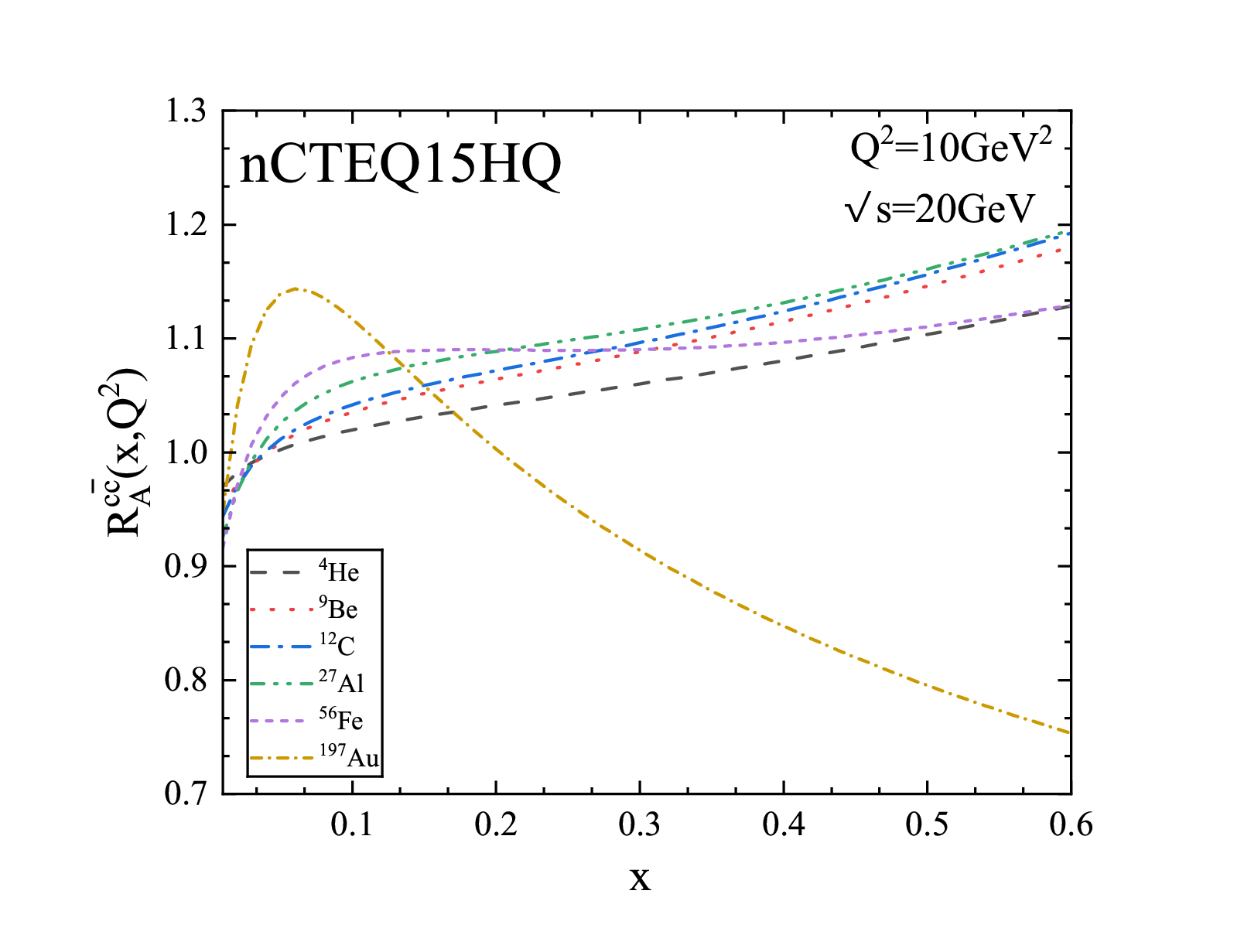}}
	\caption{$R_A^{c \bar{c}}\left(x, Q^2\right)$ defined in Eq.\,(\ref{nuclear_modi}) as a function of $x$ by using the results of global analyses from different collaborations. The typical kinematics are chosen $Q^2= 10\,\textrm{GeV}^2$, $\sqrt{s}=20\,\textrm{GeV}$. Different colors correspond to different nuclei, as indicated by the legends.}
	\label{RAcc_figure}
\end{figure*}

Next, we define a ratio factor $R_A^{c \bar{c}}$ to quantitatively assess the nuclear modification in different nuclei
\begin{eqnarray}\label{nuclear_modi}
  R_A^{c \bar{c}}\left(x, Q^2\right)=\frac{\sigma_{A,red}^{c \bar{c}}\left(x, Q^2\right)}{A \sigma_{N,red}^{c \bar{c}}\left(x, Q^2\right)},
\end{eqnarray}
where $\sigma_{N,red}^{c \bar{c}}\left(x, Q^2\right)$ in the denominator denotes the reduced cross section in electron-proton (considered free) collision. By utilizing the global analyses of the EPPS21, nNNPDF3.0(no LHCb D data), TUJU21, and nCTEQ15HQ collaborations, we can present this ratio factor $R_A^{c \bar{c}}$ as a function of $x$ for nuclei including $^{3}\mathrm{He}$, $^{4}\mathrm{He}$, $^{9}\mathrm{Be}$, $^{12}\mathrm{C}$, $^{27}\mathrm{Al}$, $^{56}\mathrm{Fe}$, $^{131}\mathrm{Xe}$ and $^{197}\mathrm{Au}$, as illustrated in Fig.\,\ref{RAcc_figure}. Here, with $Q^2= 10\,\textrm{GeV}^2$ and $\sqrt{s}=20\,\textrm{GeV}$, we are considering the kinematics expected at the forthcoming EIC \cite{AbdulKhalek:2021gbh}. It is noteworthy that the EPPS21 analysis uses the CT18ANLO set as its free proton baseline \cite{Hou:2019qau}, whereas the other three analyses provide free proton PDFs determined within their own respective global analyses frameworks.  

It can be observed that the results from all four groups exhibit ``EMC-like'' behavior in terms of gluons. While the shapes vary, this  ratio factor $R_A^{c \bar{c}}$ consistently decreases for different nuclei in the intermediate $x$-region. In the nCTEQ15HQ results, the behaviors of light nuclei and heavy nuclei are distinctly different. For light nuclei, this ratio steadily increases. This significant discrepancy shown in nCTEQ15HQ compared to other groups indicates our incomplete understanding of gluon nPDFs and suggests a need for reassessment of the parametrization in nCTEQ15HQ.

The magnitude of the gluon EMC effect for nucleus $A$ can be quantified by fitting the slope of the reduced cross section ratio $R_A^{c \bar{c}}$, i.e., $(-dR_A^{c \bar{c}}/dx)$. Based on Fig.\,\ref{RAcc_figure}, the fitting range within the ``EMC-like" region can be determined, which is chosen as $0.1\leq x \leq 0.2$ for EPPS21, and $0.3\leq x \leq 0.4$ for nNNPDF3.0(no LHCb D), nCTEQ15HQ, and TUJU21. We further expand the fitting range by 50\% for each, serving as a source of systematic uncertainty in our analysis. The obtained results are collected in Tables \ref{slopeandsub_EPPS21}-\ref{slopeandsub_nCTEQ15HQ} in App.\,\ref{appendixA}.

\begin{figure*}[htbp]
  \centering
  \subfigbottomskip=2pt
  \subfigcapskip=-5pt
  \subfigure[EPPS21]{
  \label{g23D_EPPS21}
  \includegraphics[width=0.43\linewidth]{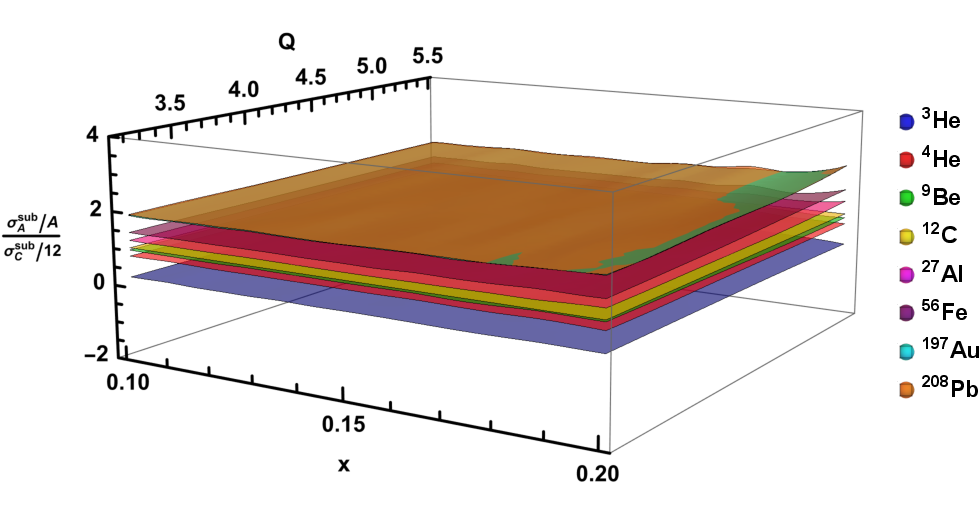}}
  \quad
  \subfigure[nNNPDF3.0(no LHCb D)]{
  \label{g23D_nNNPDF30}
  \includegraphics[width=0.43\linewidth]{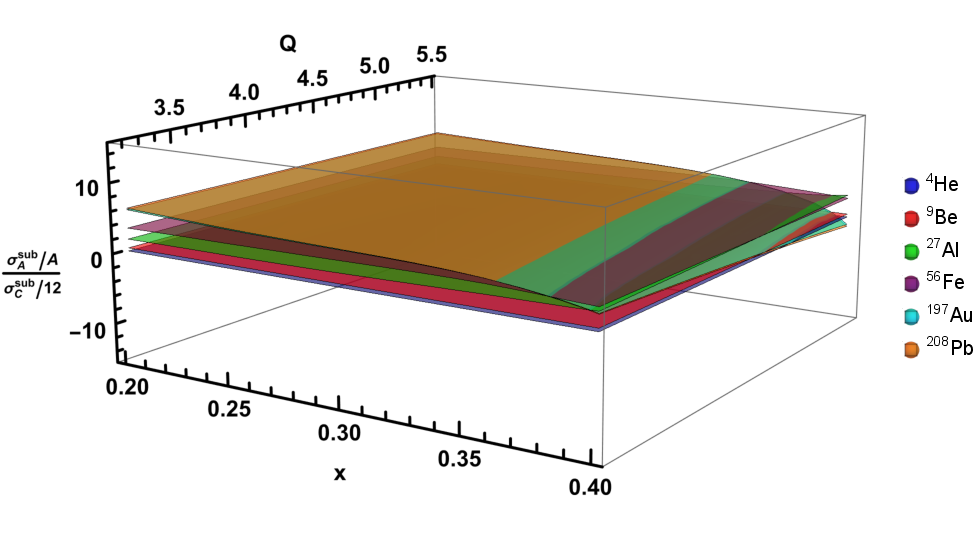}}
		
  \subfigure[TUJU21]{
  \label{g23D_TUJU21}
  \includegraphics[width=0.43\linewidth]{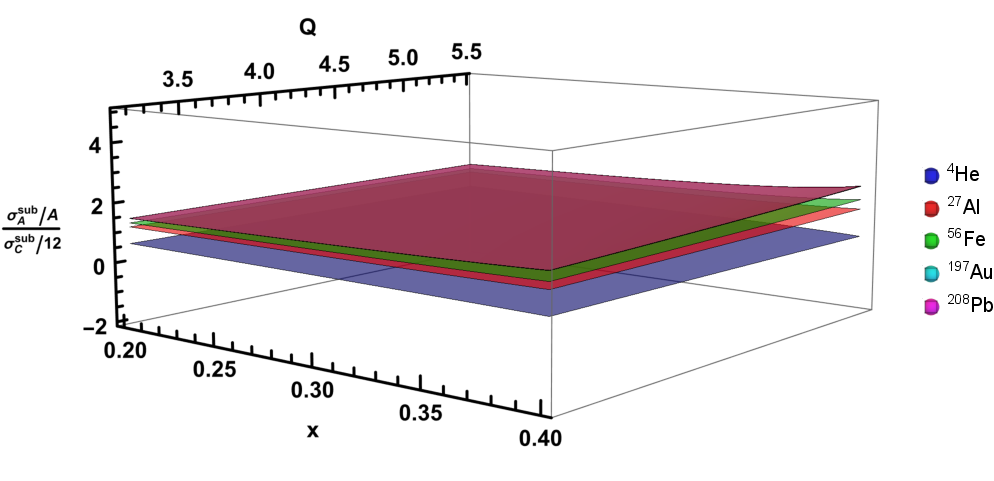}}
  \quad
  \subfigure[nCTEQ15HQ]{
  \label{g23D_nCTEQ}
  \includegraphics[width=0.43\linewidth]{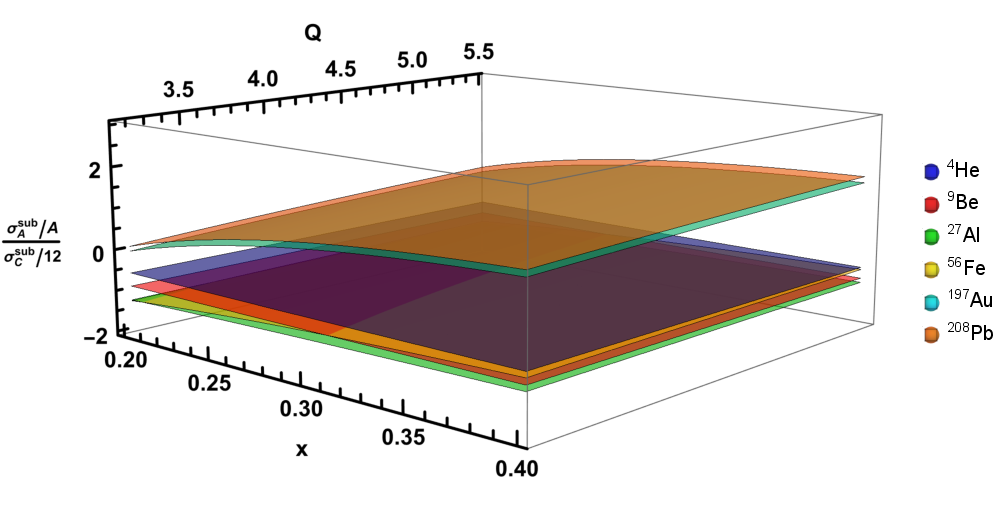}}
  \caption{The ratio $\frac{\sigma_{A}^{sub}\!/A}{\sigma_{A'}^{sub}\!/A'}$ in Eq.\,(\ref{ratioofsubcross}) for different nuclei with respect to Carbon, at different $x$ and $Q^2$.}
  \label{g23D_figure}
\end{figure*}

The authors in Ref.\,\cite{Wang:2024cpx} applied the chiral effective field theory ($\chi$EFT) and proposed a linear relation between the slope of reduced cross section ratios and the $J/\psi$ photoproduction cross section at the photon energy $E_\gamma = 7.0\,\textrm{GeV}$ in the nucleus rest frame,
\begin{eqnarray}\label{LinearRelationEFT2}
 -\frac{dR_A^{c \bar{c}}(x,Q^2)}{d x} = \left. C'(x,Q^2) \, (\sigma_{A}^{sub}\!/A) \right|_{E_{\gamma}\sim 7\,\textrm{GeV}}  \,,
\end{eqnarray}
here the function $C'(x,Q^2)$ does not depend on the type of nucleus $A$. In addition, the ratio of cross sections is linked to the ratio of gluon nPDFs by the following formula \cite{Wang:2024cpx},
\begin{eqnarray}\label{ratioofsubcross}
  \left. \frac{\sigma_{A}^{sub}\!/A}{\sigma_{A'}^{sub}\!/A'} \right|_{E_{\gamma}\sim 7\,\textrm{GeV}} \!\simeq\!  \frac{\Big(f_g^A(x, Q^2)/f_g^N(x, Q^2)\Big)-1}{\Big(f_g^{A'}(x, Q^2)/f_g^N(x, Q^2)\Big)-1} \,,
\end{eqnarray}
where $f_g^N(x, Q^2)$ is the gluon PDF in a free nucleon. It has been argued that this ratio is independent on $x$ and $Q^2$, indicating these dependencies cancel out on the right-hand side of Eq.\,(\ref{ratioofsubcross}). Here, we corroborate this characteristic by presenting three-dimensional visualizations
of this ratio. The nucleus $A'$ in the denominator is chosen to be $^{12}$C. One can observe that in all the subplots in Fig.\,\ref{g23D_figure}, this ratio displays a plateau with small variations. The value of the ratio can be determined by fitting the heights of these plateaus. Currently, there is very little experimental data available for sub-threshold production of $J/\psi$. We therefore use the calculated per-nucleon cross section in $\gamma ^{12}$C collision $(\sigma_{\textrm{C}}^{sub}\!/12) = 14.37\,\textrm{pb}$ at the photon energy $E_\gamma \sim 7\,\textrm{GeV}$ in a recent theoretical work \cite{Wang:2024cpx}. Therefore, the per-nucleon cross section for different nuclei can be extrapolated. The quantitative outcomes are presented in Tables \ref{slopeandsub_EPPS21}-\ref{slopeandsub_nCTEQ15HQ} in App.\,\ref{appendixA}.

\begin{figure*}[htbp]
  \centering
  \subfigure[EPPS21]{
  \label{sigma2_EPPS21}
  \includegraphics[width=0.43\linewidth, trim=20 20 20 20, clip]{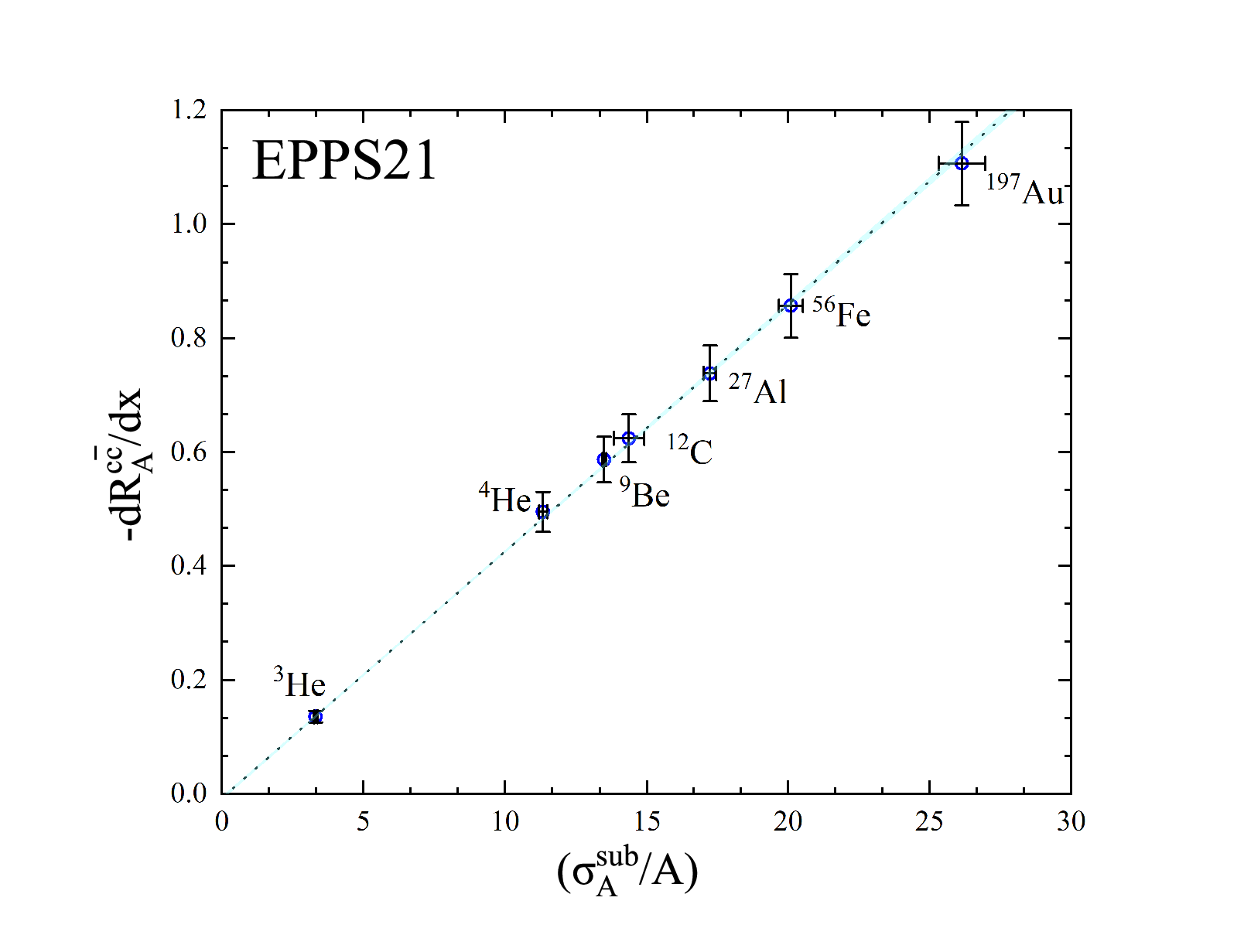}}
    \hspace{-20pt}
    \subfigure[nNNPDF3.0(no LHCb D)]{
    \label{sigma2_nNNPDF30}
  \includegraphics[width=0.43\linewidth, trim=20 20 20 20, clip]{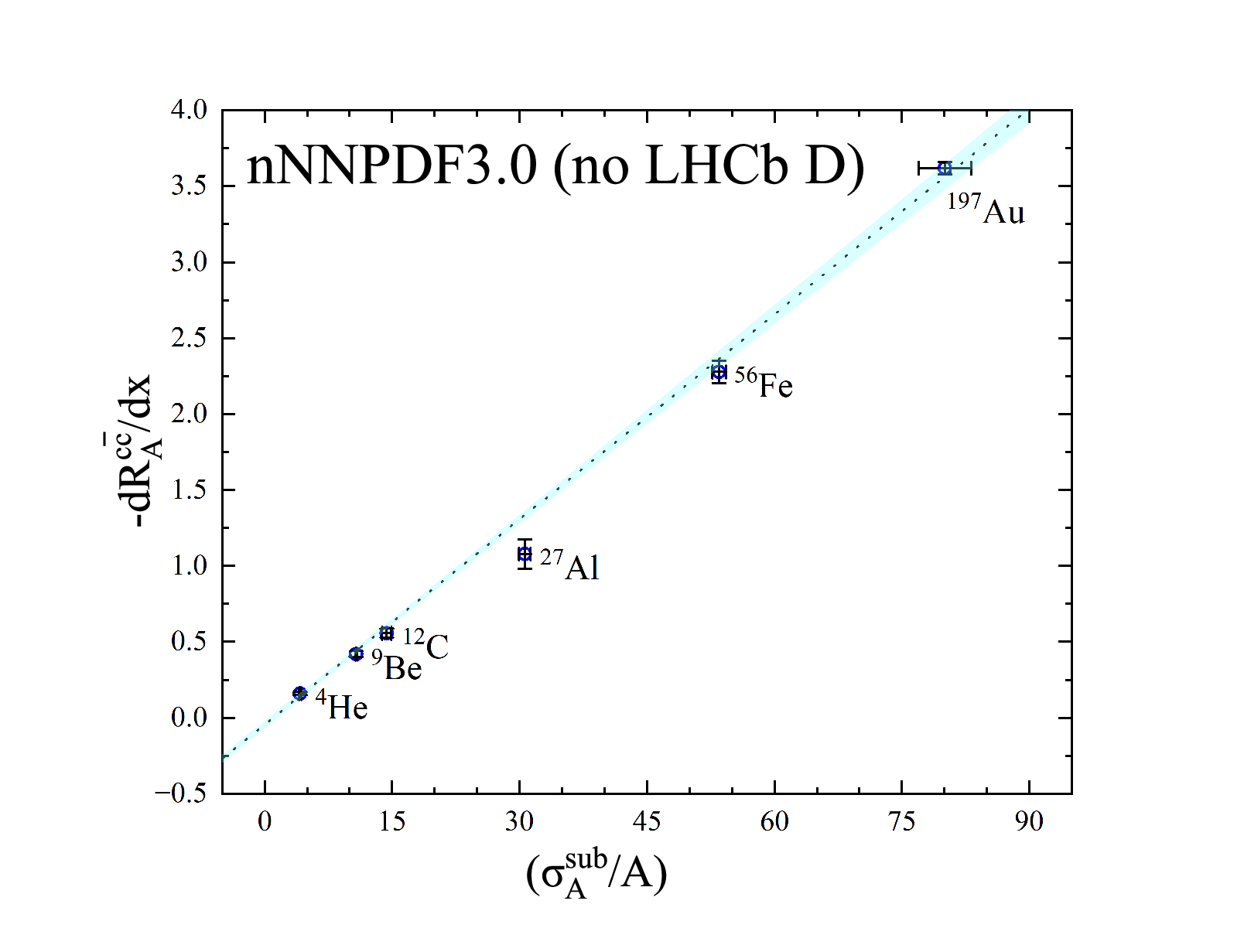}}
  \vspace{-8pt}
    \subfigure[TUJU21]{
	\label{sigma2_TUJU21}
  \includegraphics[width=0.43\linewidth, trim=20 20 20 20, clip]{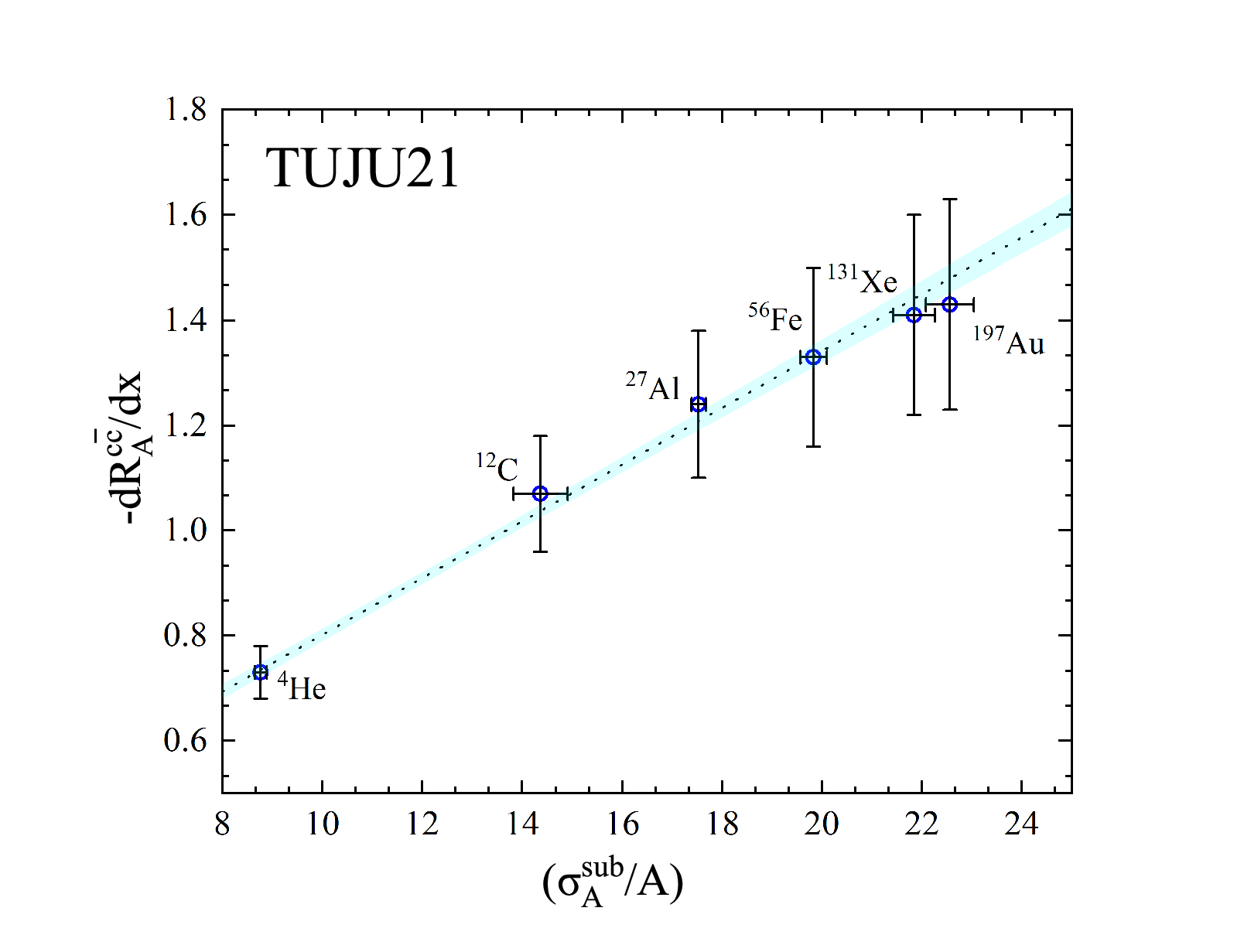}}
    \hspace{-10pt}
    \subfigure[nCTEQ15HQ]{
	\label{sigma2_nCTEQ}
  \includegraphics[width=0.43\linewidth, trim=20 20 20 20, clip]{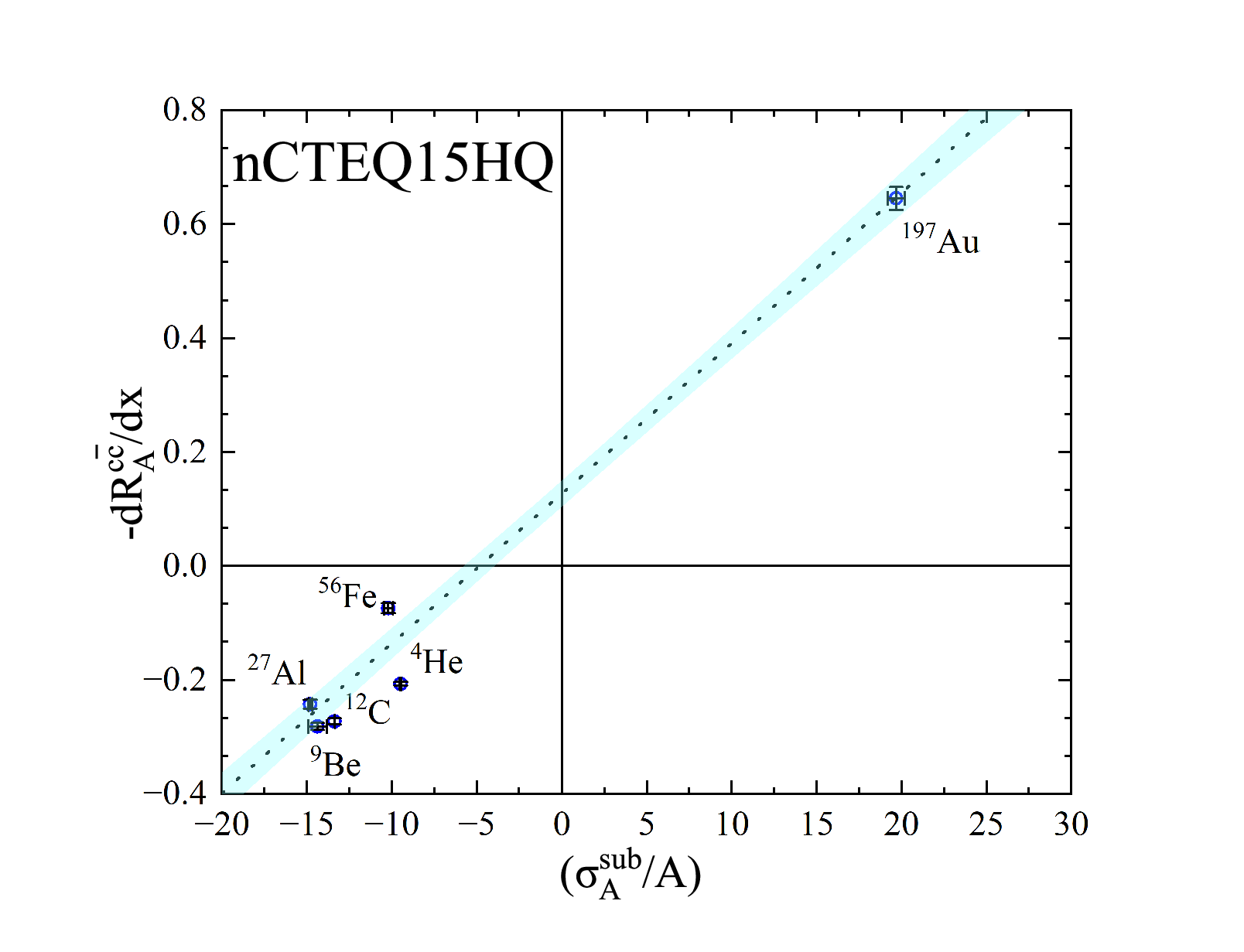}}
    \vspace{10pt}
  \caption{The linear relation between the slope of nuclear modification $(-dR_A^{c \bar{c}}/dx)$ and the sub-threshold cross section $(\sigma_{A}^{sub}\!/A)$. The gluon nPDFs are adopted from different collaborations (EPPS21, nNNPDF3.0(no LHCb D), TUJU21, and nCTEQ15HQ, respectively). The black dashed lines correspond to fits of numerical results.}
  \label{LinearRelation2}
\end{figure*}

\begin{widetext}
Analyzing the information delineated in these tables, we plot the gluon EMC slopes versus the cross sections for sub-threshold $J/\psi$ production in Fig.\,\ref{LinearRelation2}. One can find that a linear correlation exists between $(-dR_A^{c \bar{c}}/dx)$ and $(\sigma_{A}^{sub}\!/A)$. Specifically, we have
\begin{subequations}\label{LinearFit}
\begin{eqnarray}
  &&\textrm{\textbf{EPPS21:}} \nn\\
  &&-d R_A^{c\bar c} / d x = (0.043 \pm 0.0003) \times  (\sigma_{A}^{sub}\!/A)  -(0.007\pm0.003)  \,, \\
  &&\textrm{\textbf{nNNPDF3.0(no LHCb D):}} \nn\\
  &&-d R_A^{c\bar c} / d x = (0.045 \pm 0.001)\times (\sigma_{A}^{sub}\!/A)  -(0.043\pm0.021) \,, \\
  &&\textrm{\textbf{TUJU21:}} \nn\\
  &&-d R_A^{c\bar c} / d x = (0.054 \pm 0.002)\times  (\sigma_{A}^{sub}\!/A) +(0.262\pm0.027)  \,, \\
  &&\textrm{\textbf{nCTEQ15HQ:}} \nn\\
  &&-d R_A^{c\bar c} / d x = (0.026 \pm 0.001)\times  (\sigma_{A}^{sub}\!/A) +(0.128\pm0.020)  \,.
\end{eqnarray}
\end{subequations}
\end{widetext}
Although the fitted slopes and intercepts differ significantly, the results of the three groups EPPS21, nNNPDF3.0(no LHCb D), and TUJU21 all exhibit apparent linear relationship. The linear relationship presented by nCTEQ15HQ is less clear, but it broadly aligns with Eq.\,(\ref{LinearRelationEFT2}). In nCTEQ15HQ, the steadily increasing behavior of $R_A^{c \bar{c}}\left(x, Q^2\right)$ for light nuclei in Fig.\,\ref{RAcc_nCTEQ} results in negative sub-threshold cross sections, which are not physically meaningful. However, the compliance with the linear relation Eq.\,(\ref{LinearRelationEFT2}) shown in Fig.\,\ref{sigma2_nCTEQ} is still noteworthy. On the other hand, the negative cross sections stress the need for reassessment of the parametrization in nCTEQ15HQ.

The $\chi$EFT provides us with a method for scale separation, which naturally leads to the linear relation in Eq.\,(\ref{LinearFit}). What we want to highlight is that the scale separation itself provides valuable information, such as the existence of a linear relationship, regardless of the specific slope and intercept values. The results obtained here support the conclusions in Ref.\,\cite{Wang:2024cpx}, but also underscore our limited knowledge of gluon nPDFs, whose error are not taken into account. However, the introduction of the ratio factor $R_A^{c \bar{c}}$ would help to reduce the corresponding uncertainties substantially. Furthermore, the isoscalar correction dependencies of nPDFs are significantly less pronounced in gluons compared to quarks. At the end of this section, we wish to emphasize that Fig.\,\ref{LinearRelation2} serves as an indication, not a demonstration. Therefore, the future experimental validation (or negation) of this linear relationship is imperative.

\section{Test of linear relation of quarks}
\label{linearquark}
\begin{figure}[H]
  \centering
  \subfigure{
  \includegraphics[width=0.85\linewidth]{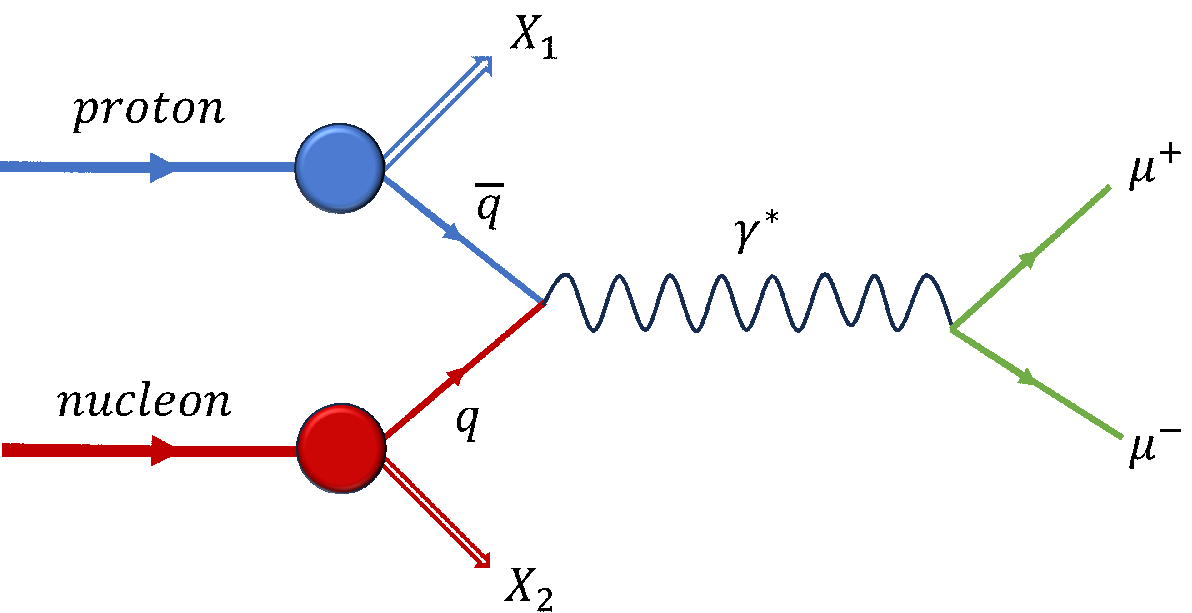}}
  \caption{The proton-induced Drell-Yan process at leading order.}
  \label{Drell_Yan_process}
\end{figure}

Apart from DIS, Drell-Yan production of lepton pairs from nucleons is another important tool to study the quark structure of nucleons \cite{Drell:1970wh,Huang:2025kmd,Chmaj:1983jq}, which inherently involves the antiquark distributions of either the beam or target hadron. The EMC effect has been experimentally studied in the  Drell-Yan reactions \cite{Moreno:1990sf,Alde:1990im,Heinrich:1989cp,AE866E789E772}. In proton-induced Drell–Yan, a quark (antiquark) with momentum fraction $x_1$ from the beam proton and an antiquark (quark) of the target nucleon with momentum fraction $x_2$ annihilate via a virtual photon into a charged-lepton pair. This process is illustrated in Fig.\,\ref{Drell_Yan_process}.
The corresponding cross section depends on the charge-squared-weighted sum of quark and antiquark distributions in the beam and the target nucleon \cite{Kulagin:2014vsa,Kenyon:1982tg},
\begin{eqnarray}\label{DrellYanFormula}
  \frac{d^2\sigma_{\textrm{DY}}(pA)}{dx_1dx_2} &=& K\frac{4\pi\alpha^2}{9Q^2}\sum_{q}e_q^2\left[f_q^p(x_1,Q^2)f_{\bar{q}}^A(x_2,Q^2) \right. \nonumber\\
  &&\quad + \left.f_{\bar{q}}^p(x_1,Q^2)f_q^A(x_2,Q^2)\right] \,.
\end{eqnarray}
The invariant mass of the lepton pair, denoted by $Q$, is given by $Q^2=s x_1 x_2$. The summation is taken over quark flavors; $\alpha$ is the fine structure constant; $e_q$ is the charge of quark or antiquark of flavor $q$; and $f_{q (\bar{q})}^{p (A)}$ corresponds to quark (antiquark) PDF in the proton (nucleus). The factor $K$ absorbs higher-order QCD corrections, which is approximately $K\sim1+(\alpha_s/2\pi)(4\pi^2/3)+\mathcal{O}(\alpha_s^2)$ \cite{Curci:1979am}.

\begin{figure*}[]
  \centering
  \subfigure[$^{12}$C]{
  \label{DY_pA_C12}
  \includegraphics[width=0.35\linewidth]{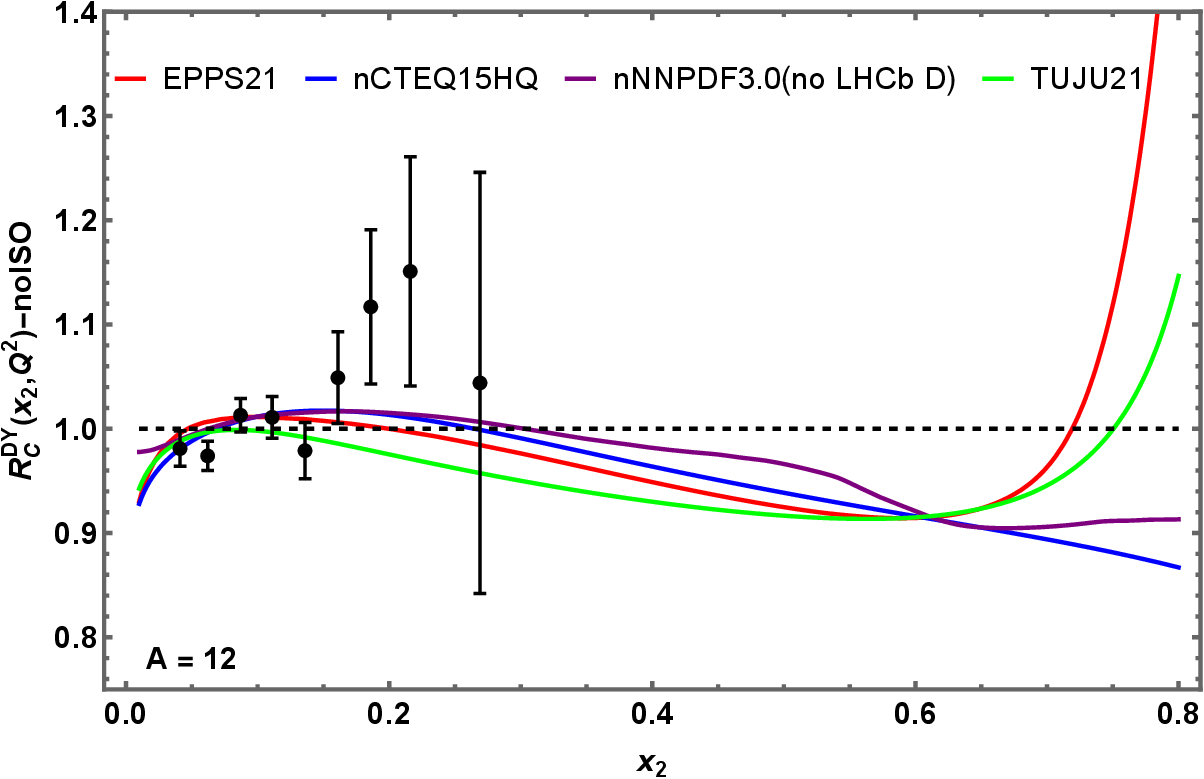}}
   \hspace{15pt}
    \subfigure[$^{40}$Ca]{
	\label{DY_pA_Ca40}
  \includegraphics[width=0.35\linewidth]{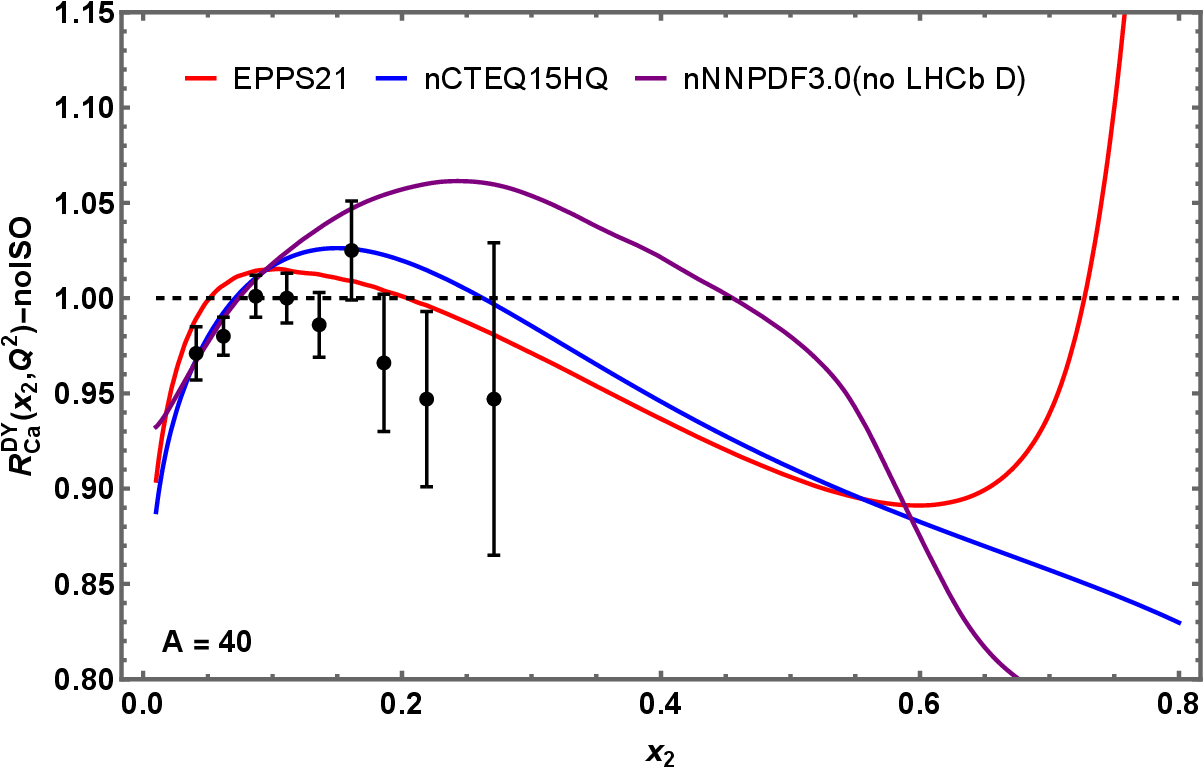}}
  
\vspace{5pt}

    \subfigure[$^{56}$Fe]{
	\label{DY_pA_Fe56}
  \includegraphics[width=0.35\linewidth]{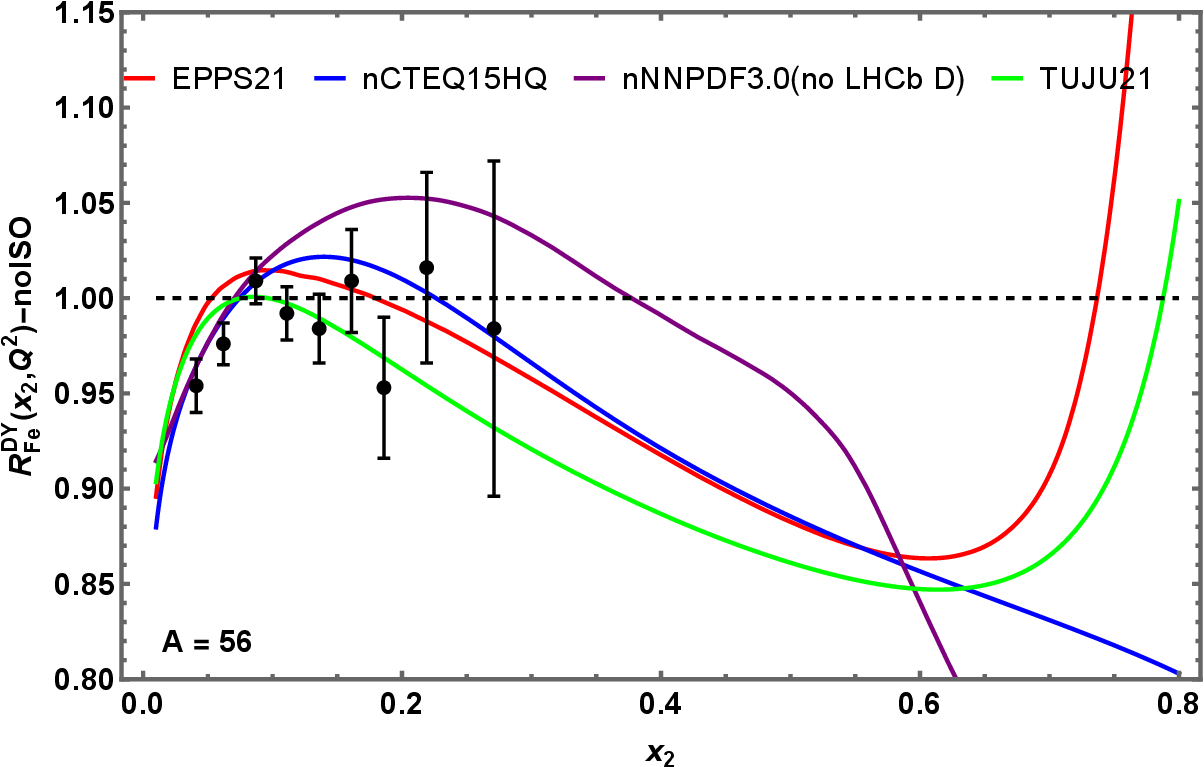}}
	\hspace{15pt}
	\subfigure[$^{184}$W]{
	\label{DY_pA_W184}
  \includegraphics[width=0.35\linewidth]{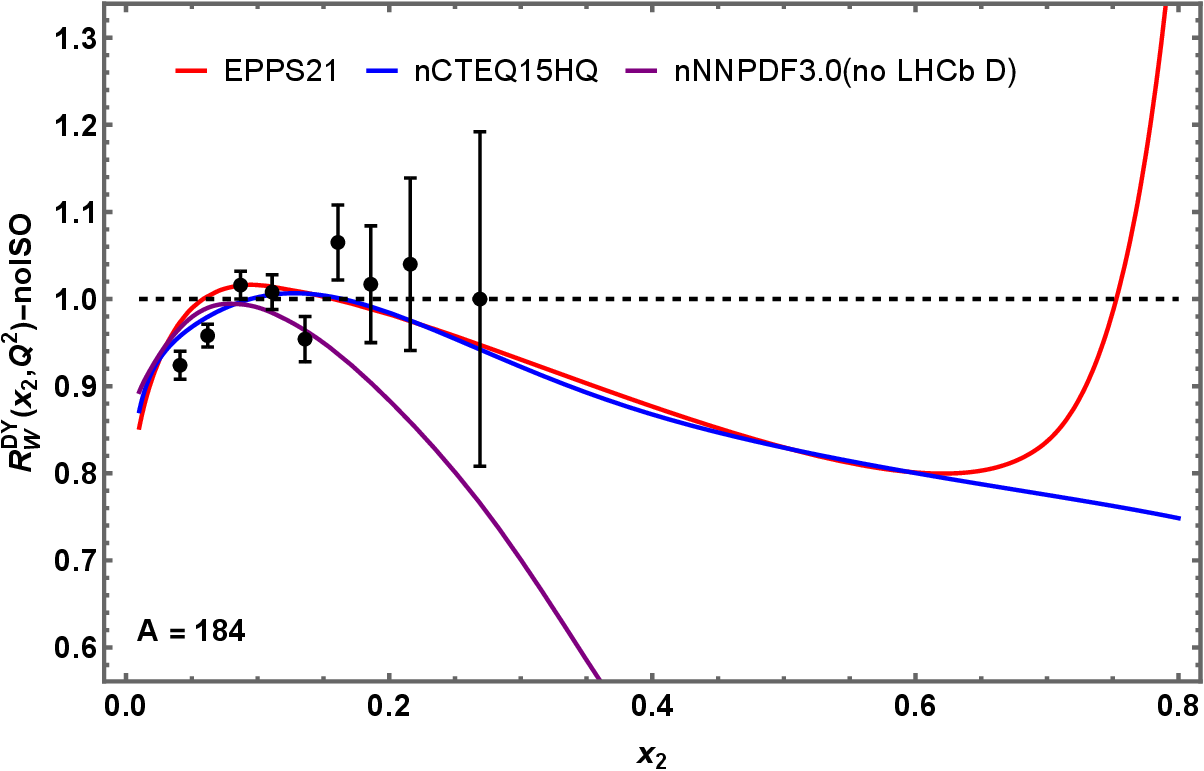}}
  \caption{The ratio of proton-induced Drell-Yan differential cross sections as a function of $x_2$ in different nuclei, the typical kinematics are chosen $Q^2= 25\,\textrm{GeV}^2$, $\sqrt{s}=40\,\textrm{GeV}$.}
  \label{DY_pA}
\end{figure*}

In Eq.\,(\ref{DrellYanFormula}), the nuclear dependence comes from the modification of (anti)quark distributions in the target nucleus. The quark distributions of nucleus $A$ are constructed as
\begin{eqnarray}
  f^A_q(x,Q^2)=\frac{Z f_q^{p/A}(x,Q^2)+(A-Z)f_q^{n/A}(x,Q^2)}{A} \,,
\end{eqnarray}
with $Z$ being the atomic charge number. Here $f_q^{p(n)/A}$ is the quark PDF of a proton (neutron) bound in the nucleus $A$. Since we are interested in the EMC effect of nucleon, we integrate the variable $x_1$ and formally obtain
\begin{align}\label{DrellYanFormula2}
	\frac{d\sigma_{\textrm{DY}}(pA)}{dx_2} &= K \frac{4\pi\alpha^2}{9Q^2}\sum_{q}e_q^2\Big[ f_{\bar{q}}^A(x_2,Q^2) \int_0^1 dx_1 f_q^p(x_1,Q^2) \nonumber \\
	&\quad + f_q^A(x_2,Q^2) \int_0^1 dx_1 f_{\bar{q}}^p(x_1,Q^2) \Big] \,.
\end{align}
Thus a ratio factor can be defined to quantify the nuclear modification in different nuclei relative to deuteron:
\begin{eqnarray}\label{nuclear_modiDY}
  R_A^{\textrm{DY}}\left(x_2, Q^2\right)=\frac{d\sigma_{\textrm{DY}}(pA)/dx_2}{d\sigma_{\textrm{DY}}(pD)/dx_2} \,.
\end{eqnarray}
We can depict this $R_A^{\textrm{DY}}$ by utilizing nPDFs from EPPS21, nNNPDF3.0(no LHCb D), TUJU21, and nCTEQ15HQ. Fig.\,\ref{DY_pA} shows a comparison of these results with the data from the Fermilab E772 experiment for a number of nuclear targets \cite{Alde:1990im,AE866E789E772}.  In the existing experimental data, it is difficult for $x_2$ to surpass 0.4, since the absolute cross section would become very small. We set $Q^2=25\,\textrm{GeV}^2$ and $\sqrt{s}=40\,\textrm{GeV}$, consistent with the kinematics of typical Drell-Yan experiments such as E772 and E866 \cite{Alde:1990im,NuSea:1999egr}. The kinematic constraint $x_1 x_2 = Q^2/s$ yields a lower integration limit for $x_{1,\textrm{min}} \approx 0.0156$ with $x_{2,\textrm{max}}=1$. Thus, the numerical integration is performed over $x_1 \in [0.0156,1]$. From Fig.\,\ref{DY_pA}, it can be observed that the shapes of the ratio $R_A^{\textrm{DY}}$ for EPPS21, nCTEQ15HQ, and TUJU21 are closely aligned, while nNNPDF3.0(no LHCb D) shows a marked deviation from the others. These results show that the predictions of $R_A^{\textrm{DY}}$ are sensitive to the nPDF sets used as input in the calculations. 

\begin{figure*}[]
	\centering
	\subfigure[EPPS12]{
		\label{DY_term12_EPPS12}
		\includegraphics[width=0.35\linewidth]{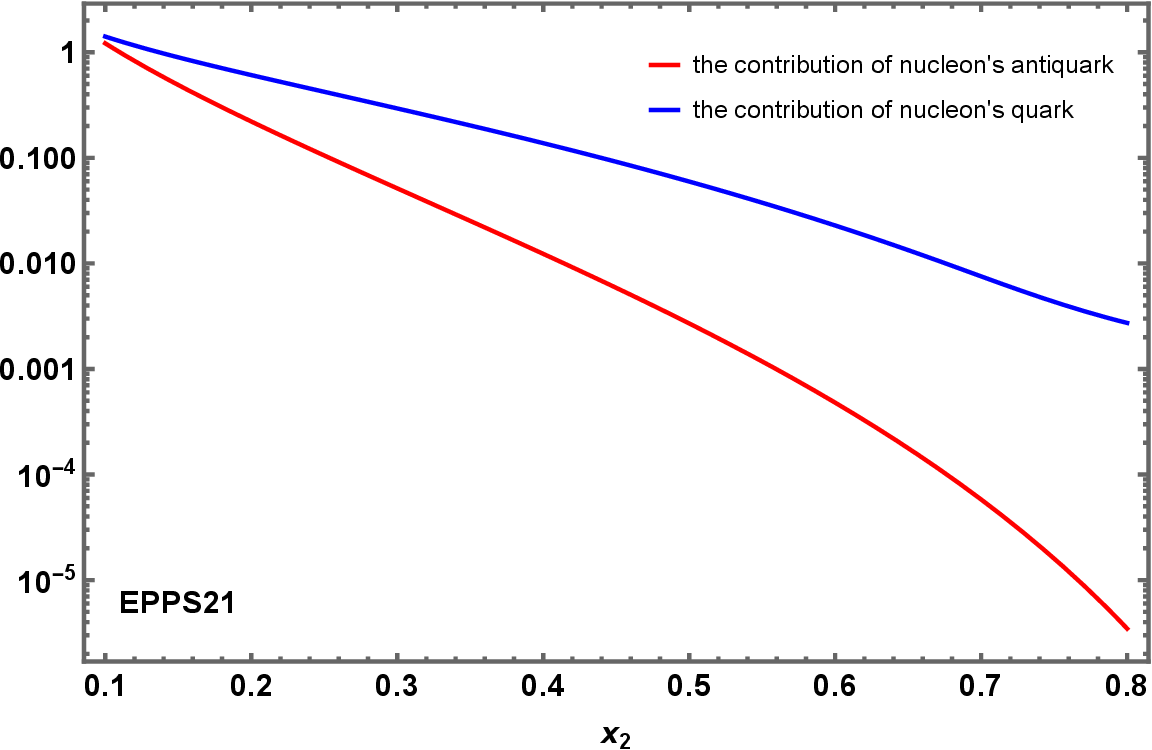}}
		\hspace{15pt}
	\subfigure[nNNPDF3.0(no LHCb D)]{
	\label{DY_term12_nNNPDF}
	\includegraphics[width=0.35\linewidth]{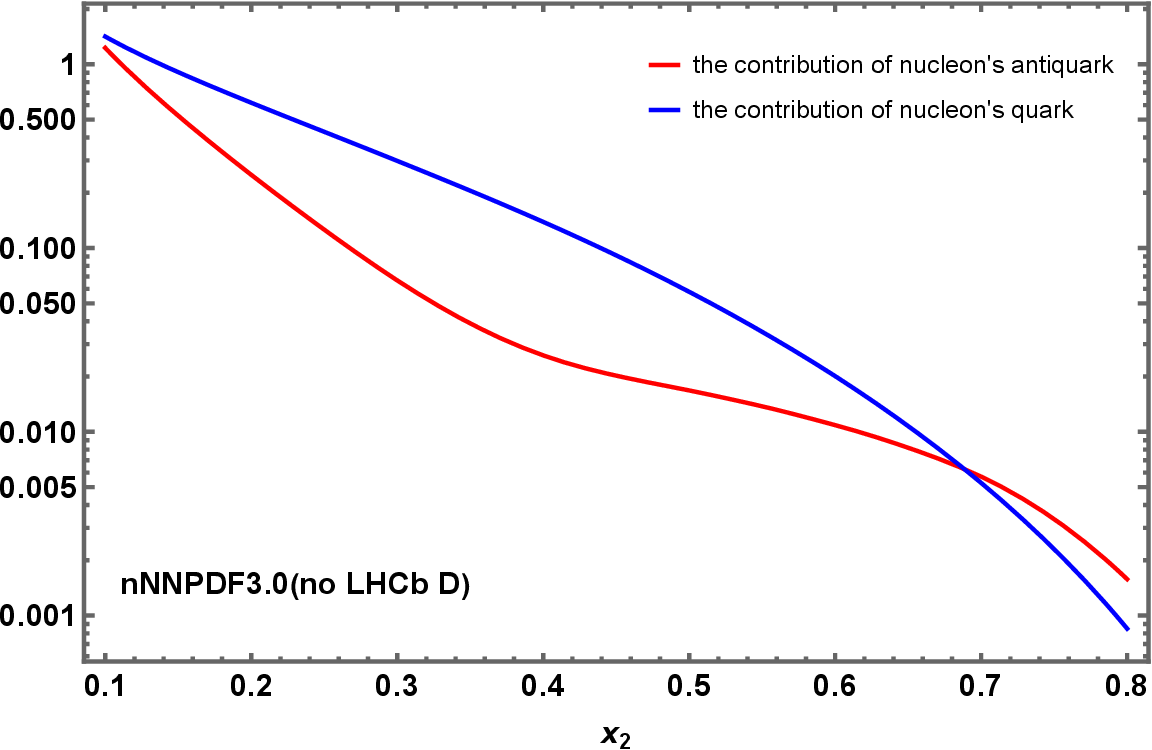}}
	
	\vspace{5pt}
	
	\subfigure[TUJU21]{
		\label{DY_term12_TUJU21}
		\includegraphics[width=0.35\linewidth]{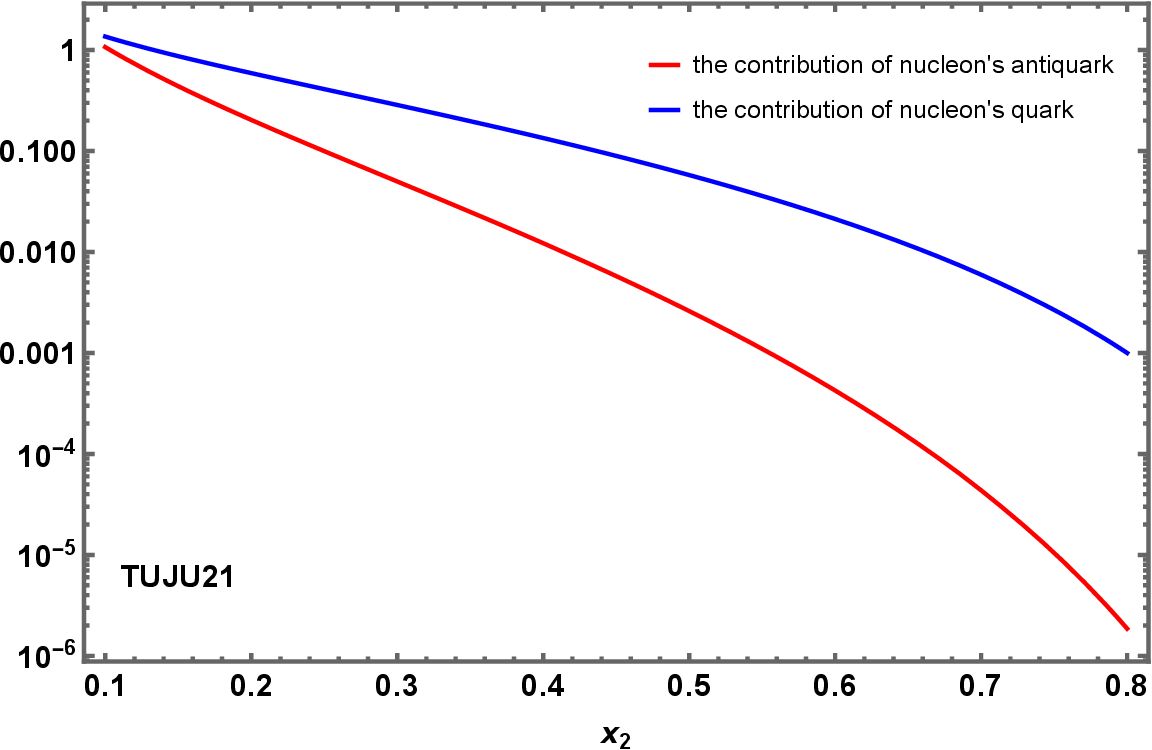}}
		\hspace{15pt}
	\subfigure[nCTEQ15HQ]{
    \label{DY_term12_nCTEQ}
	\includegraphics[width=0.35\linewidth]{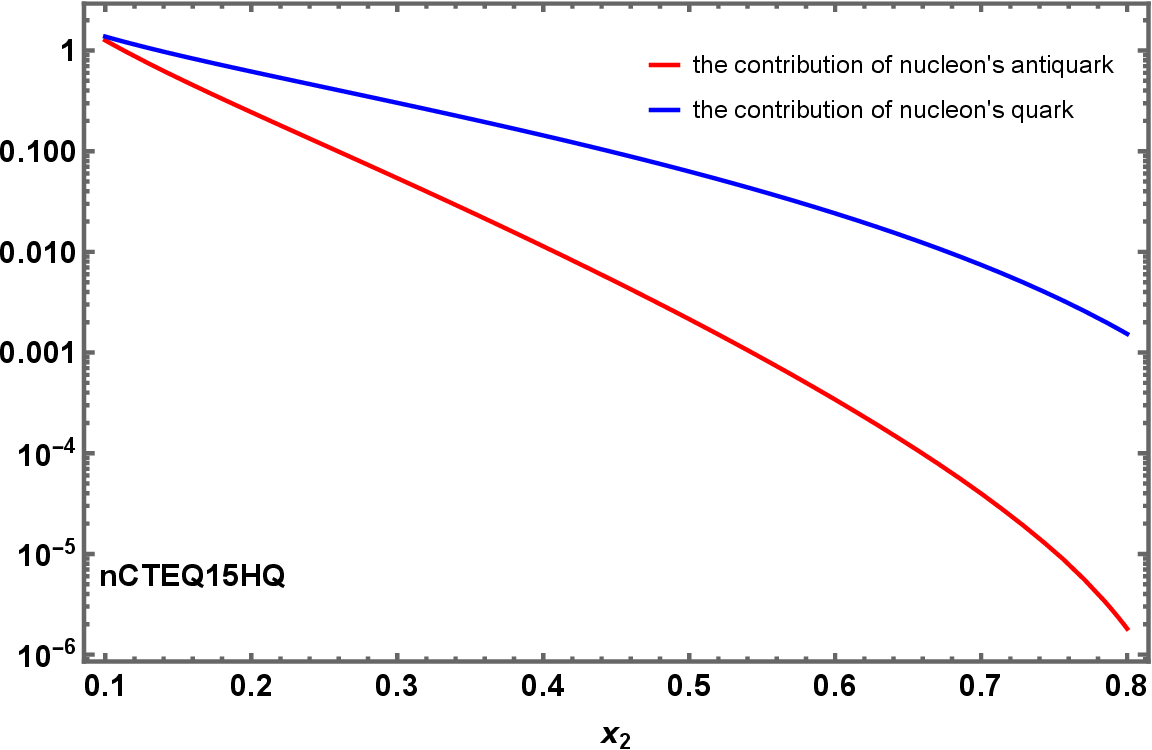}}
	\caption{Comparison between the contributions from the antiquark and quark distributions of target nucleon ($^{56}\textrm{Fe}$) in Eq.\,(\ref{DrellYanFormula}). The typical kinematics are chosen $Q^2= 25\,\textrm{GeV}^2$.}
	\label{DY_term12}
\end{figure*}

It is important to emphasize that the proton-induced Drell–Yan process involves the quark and antiquark distributions of both the beam proton and the target nucleon. Given our focus on the nucleon's substructure, we attribute the two terms in Eq.\,(\ref{DrellYanFormula2}) to contributions from the nucleon's antiquark and quark distributions, respectively. We write down these two contributions:
\begin{eqnarray}
    \left.\left\{
\begin{array}
{l}\sum_qe_q^2\left[f_{\bar{q}}^A(x_2,Q^2)\int_{x_{1,\textrm{min}}}^1dx_1 f_q^{p}(x_1,Q^2)\right], \\
 \\
\sum_qe_q^2\left[f_q^A(x_2,Q^2)\int_{x_{1,\textrm{min}}}^1dx_1 f_{\bar{q}}^{p}(x_1,Q^2)\right],
\end{array}\right.\right.
\end{eqnarray}
and compare their magnitudes in Fig.\,\ref{DY_term12}. Within the most EMC region of interest, the contribution from the target nucleon’s quark distributions is predominant. Notably, for the nNNPDF3.0(no LHCb D) case, the nucleon's quark contribution becomes smaller than the antiquark's when $x_2 \gtrsim 0.7$, a behavior that diverges from other global analyses. As noted in Ref.\,\cite{AbdulKhalek:2022fyi}, this arises because experimental constraints on large-$x_2$ nuclear antiquarks are limited, causing the methodological assumptions in the fit to play a more significant role. By restricting $x_2$ to the interval $x_2\in[0.2, 0.6]$, the contribution from nucleon's antiquark distributions can be safely removed,

\begin{align}\label{DrellYanFormula3}
	\frac{d\sigma_{\textrm{DY}}(pA)}{dx_2} &= K \frac{4\pi\alpha^2}{9Q^2}  \nn\\
	&\!\!\!\!\!\!\times\left[\sum_{q}e_q^2\Big(f_q^A(x_2,Q^2) \int_{x_{1,\textrm{min}}}^1 dx_1 f_{\bar{q}}^p(x_1,Q^2) \Big) \right]\,.
\end{align}
 The observation of EMC effects in the medium-$x_2$ region is plausible. The magnitude of EMC effect in different nuclei can be quantified by fitting the slope of $R_A^{\textrm{DY}}$, just like what we did to $R_A^{c \bar{c}}$ in the previous section.
 
\begin{figure*}[]
	\centering
	\subfigure[EPPS21]{
		\label{RDY_EPPS21}
		\includegraphics[width=0.40\linewidth, trim=20 20 20 20, clip]{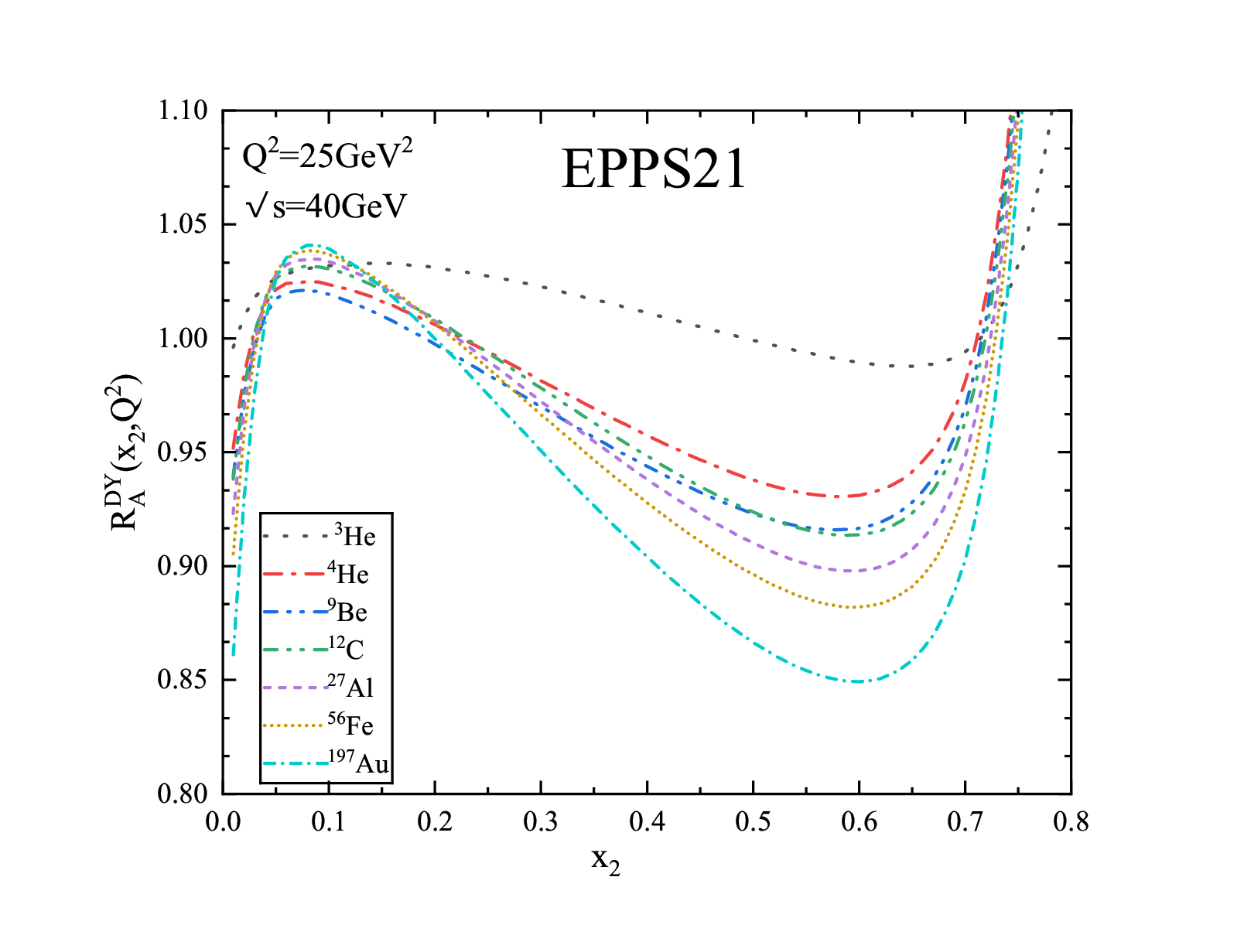}}
	\hspace{-20pt}
	\subfigure[nNNPDF3.0(no LHCb D)]{
		\label{RDY_nNNPDF30}
		\includegraphics[width=0.40\linewidth, trim=20 20 20 20, clip]{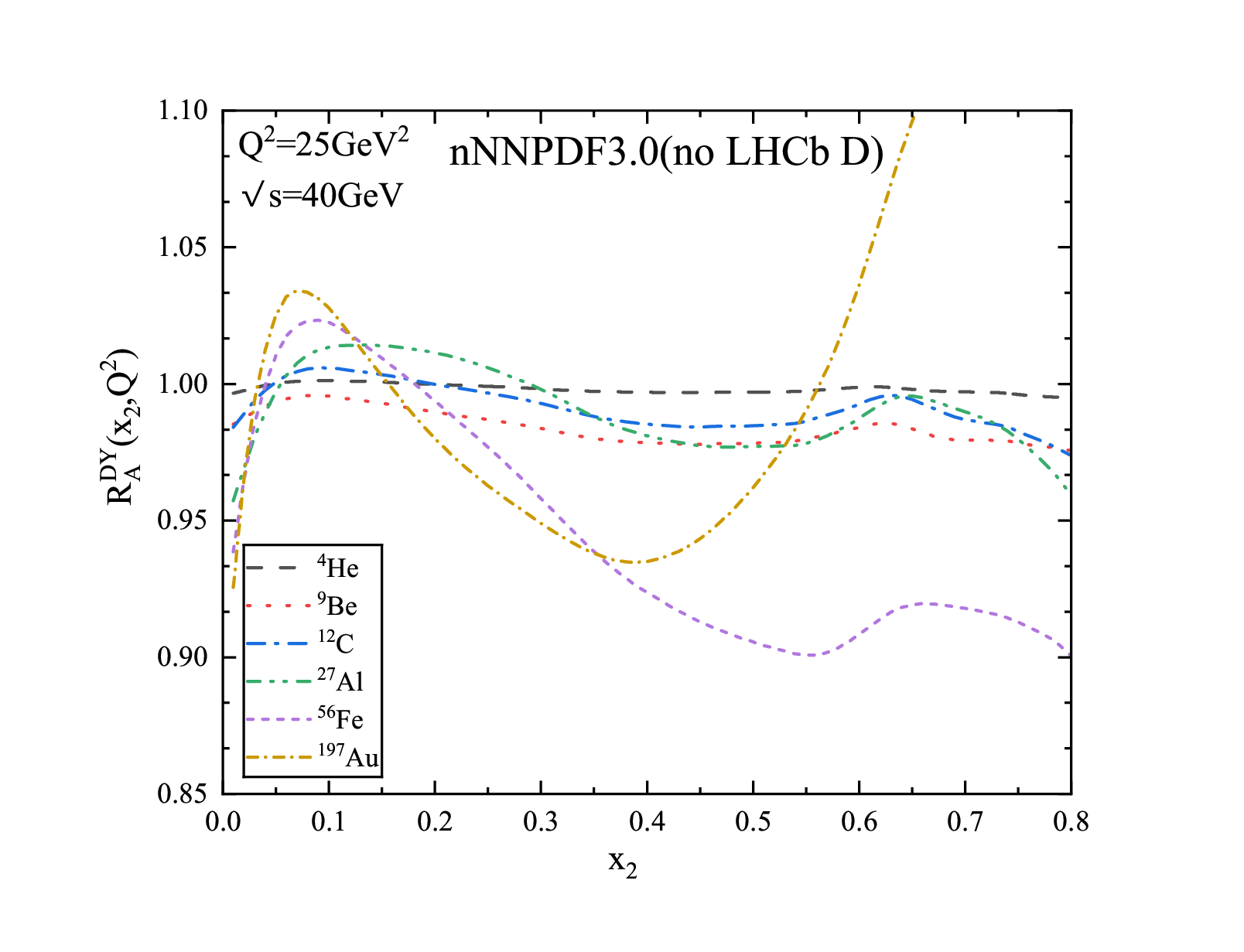}}
	\vspace{-10pt}
	\subfigure[TUJU21]{
		\label{RDY_TUJU21}
		\includegraphics[width=0.40\linewidth, trim=20 20 20 20, clip]{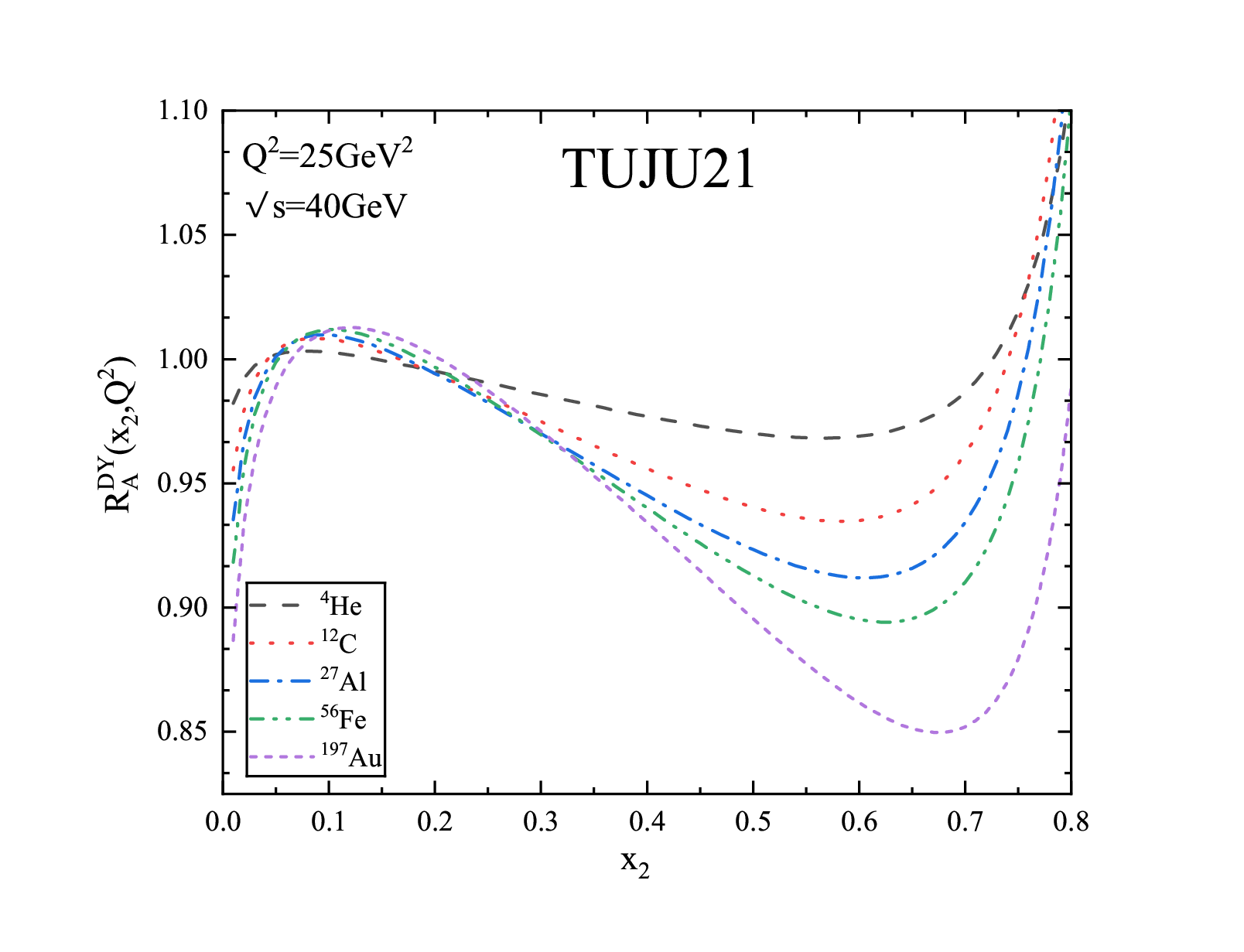}}
	\hspace{-20pt}
	\subfigure[nCTEQ15HQ]{
		\label{RDY_nCTEQ}
		\includegraphics[width=0.40\linewidth, trim=20 10 20 20, clip]{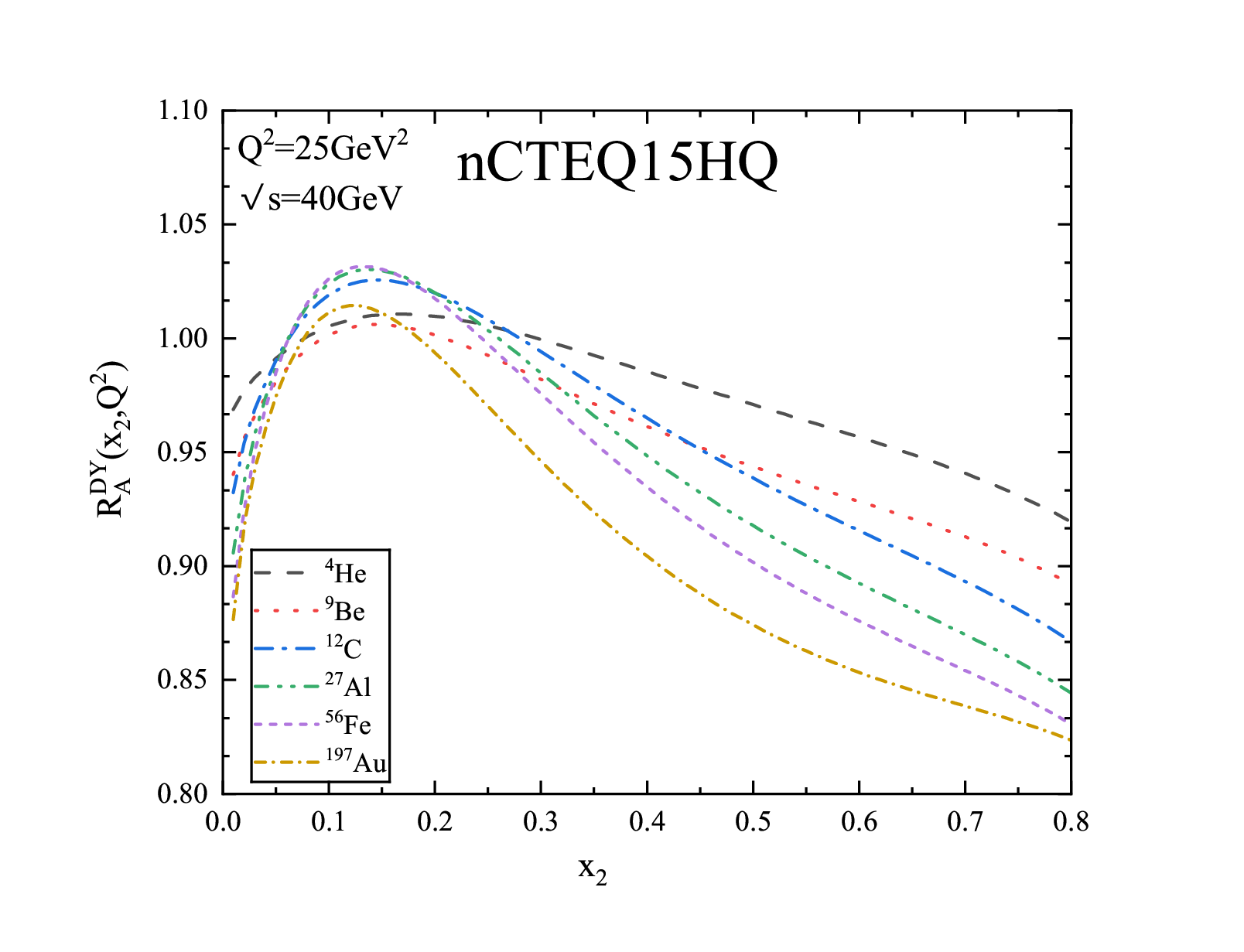}}
		 \vspace{10pt}
	\caption{ $R_A^{\textrm{DY}}\left(x_2, Q^2\right)$ defined in Eq.\,(\ref{DrellYanFormula3}) as a function of $x_2$ by using the global analyses from different groups. The typical kinematics are chosen $Q^2= 25\,\textrm{GeV}^2$, $\sqrt{s}=40\,\textrm{GeV}$. Different colors correspond to different nuclei, as indicated by the legends.}
	\label{RDY_figure}
\end{figure*}

 To account for the unequal number of protons and neutrons in certain nuclei, one can include an isoscalar correction factor (ISO) which is defined as \cite{CLAS:2019vsb}
\begin{eqnarray}
	\textrm{ISO} = \frac{\frac{A}{2}(1+\frac{\sigma_n}{\sigma_p})}{Z+N \frac{\sigma_n}{\sigma_p}} \,.
\end{eqnarray}
where $\sigma_n$ and $\sigma_p$ are the elementary electron-neutron and electron-proton cross sections, respectively. The calculation of their ratio $\sigma_n/\sigma_p$ is model-dependent, here we adopt $\sigma_n/\sigma_p \approx 1-0.8 x_2$  \cite{Hen:2014nza}. This correction factor adjusts the per-nucleon cross section for nucleus $A$ to a new value which represents the per-nucleon cross section for a ``virtual nucleus'' $A$ with equal numbers of neutrons and protons. We present in Fig.\,\ref{RDY_figure} the ratio $R_A^{\textrm{DY}}$ computed using the ISO factor for four different nPDF sets. The results from EPPS21 and TUJU21 are generally consistent with each other, whereas those from nCTEQ15HQ and especially nNNPDF3.0(no LHCb D) show distinct behaviors.

\begin{figure*}[]
  \centering
  \subfigure[EPPS21]{
  \label{linearDY_EPPS21}
  \includegraphics[width=0.4\linewidth, trim=20 20 20 20, clip]{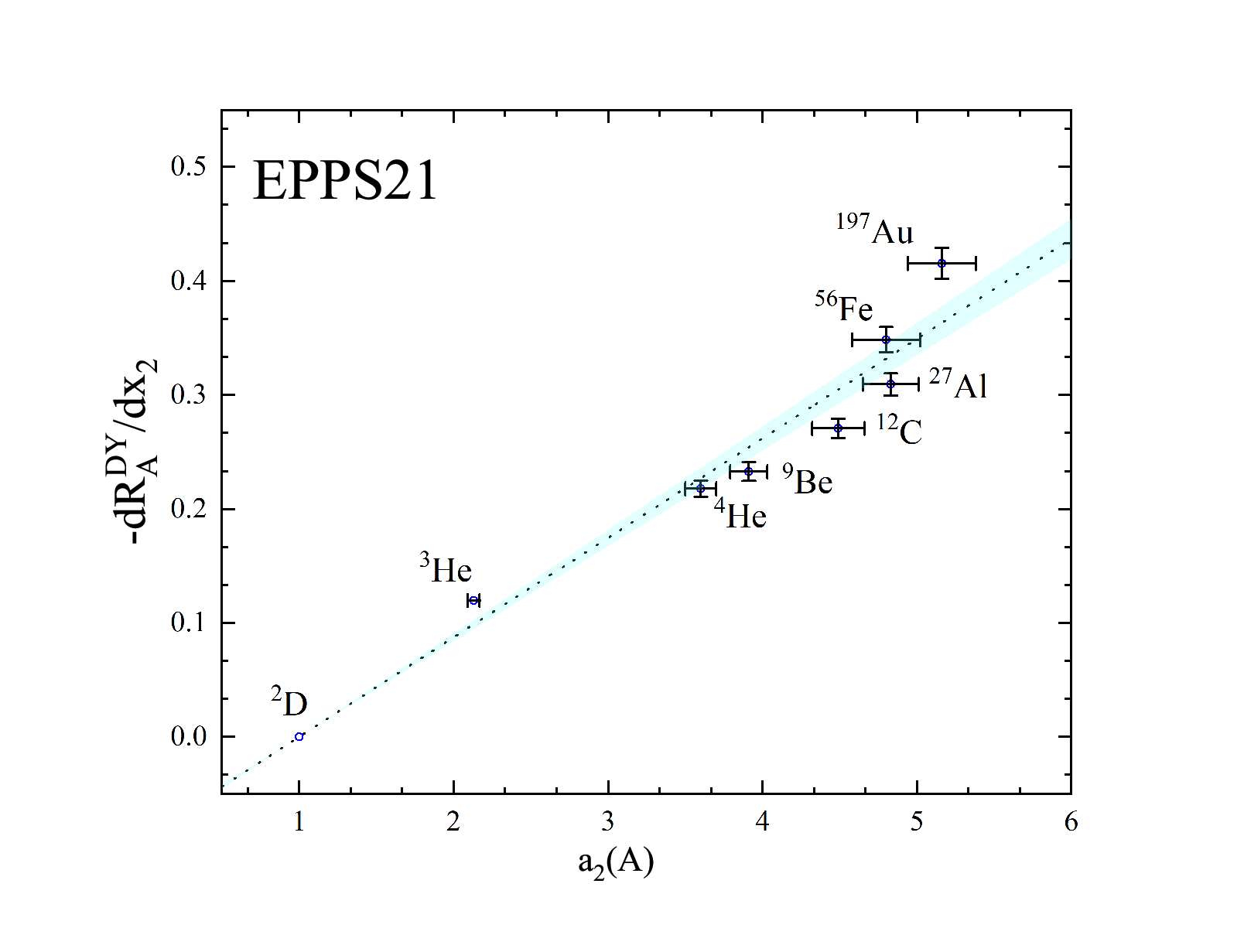}}
    \hspace{-20pt}
    \subfigure[nNNPDF3.0(no LHCb D)]{
	\label{linearDY_nNNPDF30}
  \includegraphics[width=0.4\linewidth, trim=20 20 20 20, clip]{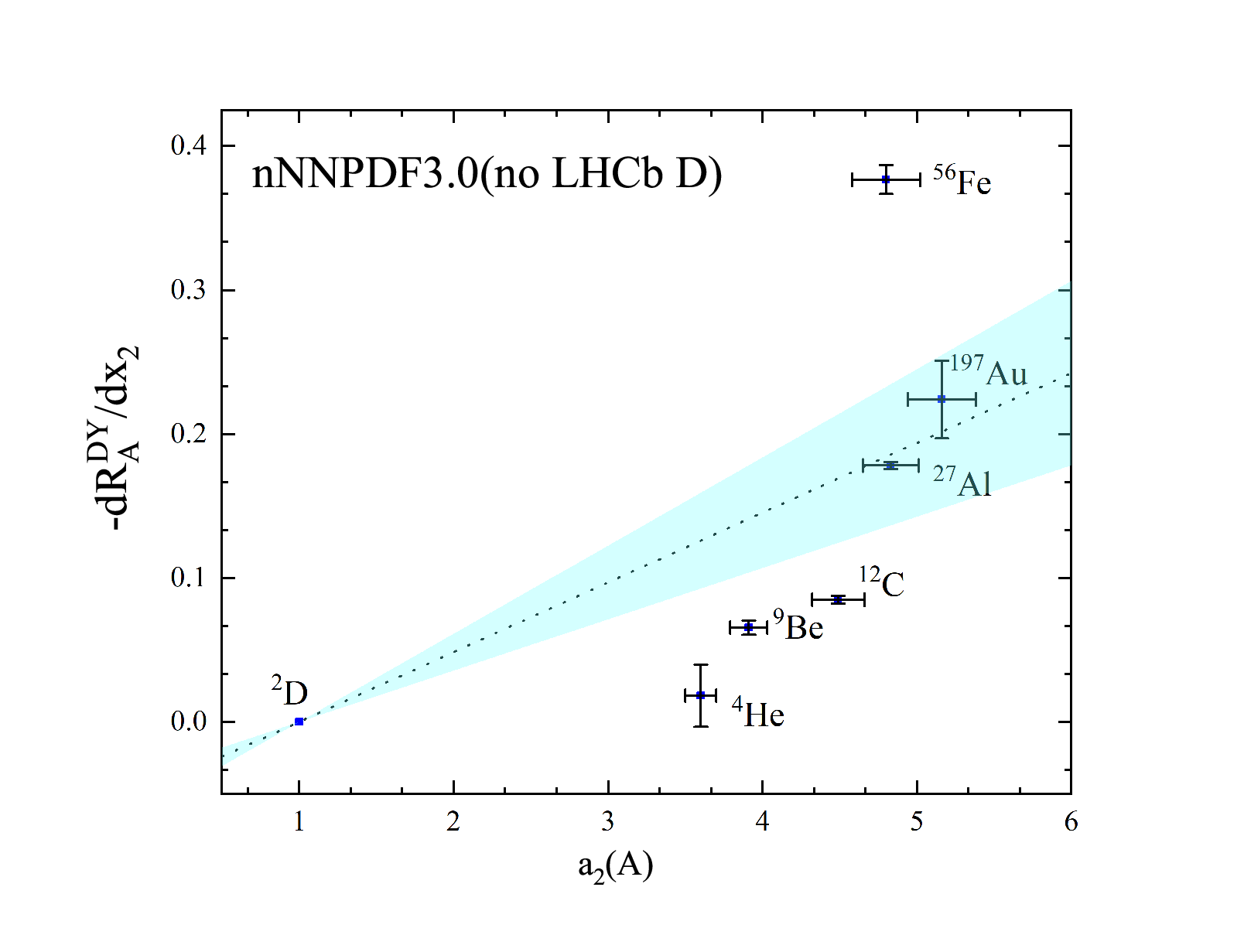}}
	\vspace{-10pt}
    \subfigure[TUJU21]{
	\label{linearDY_TUJU21}
  \includegraphics[width=0.4\linewidth, trim=20 20 20 20, clip]{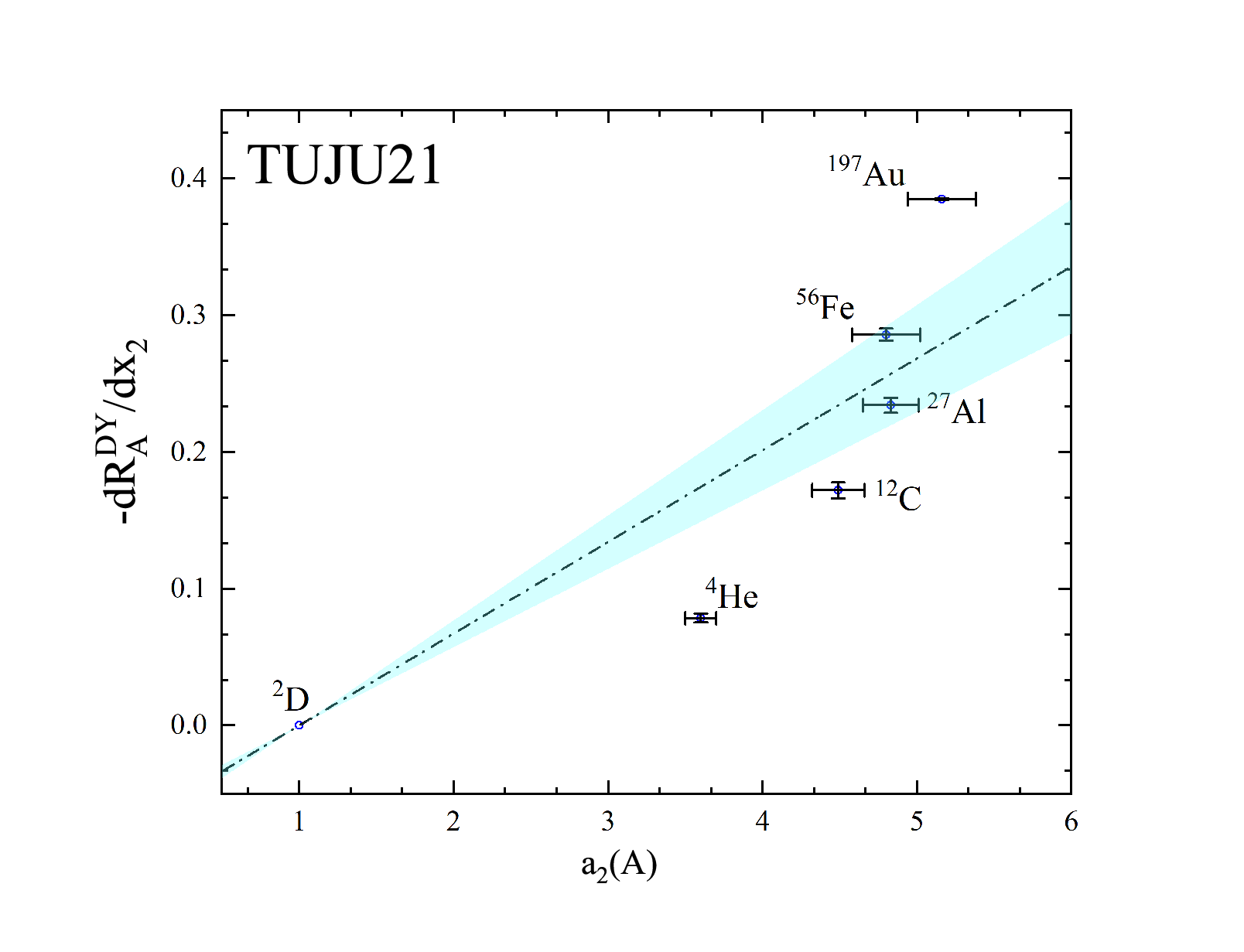}}
	\hspace{-20pt}
	\subfigure[nCTEQ15HQ]{
	\label{linearDY_nCTEQ}
  \includegraphics[width=0.4\linewidth, trim=20 20 20 20, clip]{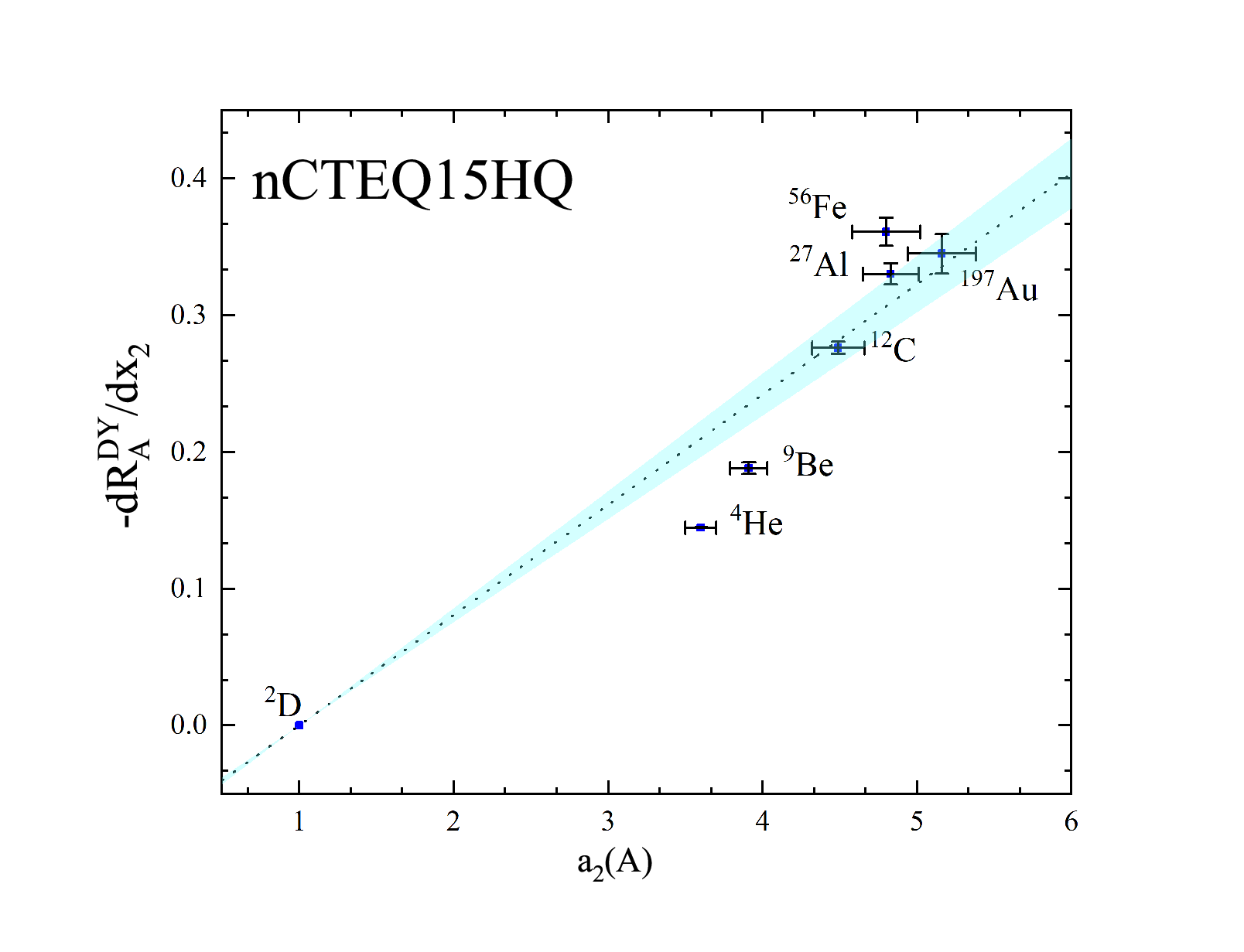}}
    \vspace{10pt}
  \caption{ The linear relation between the slope of $R_A^{\textrm{DY}}(x_2, Q^2)$ and the SRC scaling factor $a_2(A)$. The quark nPDFs are adopted from different groups (EPPS21, nNNPDF3.0(no LHCb D), TUJU21, and nCTEQ15HQ, respectively). The black dashed lines correspond to fits of numerical results, which is normalized to a reference point with $-d R_A^{\textrm{DY}}(D)/dx_2=0$ and $a_2(D)=1$.}
  \label{linearDY_figure}
\end{figure*}

The results for the slopes of $R_A^{\textrm{DY}}$ in the EMC region are collected in Tables \ref{slopeandSRC_EPPS21_DY}-\ref{slopeandSRC_nCTEQ15HQ_DY} in App.\,\ref{appendixB}. The fitting range is $0.35\leq x_2 \leq 0.45$ for the nPDFs of EPPS21, nCTEQ15HQ, and TUJU21. For nNNPDF3.0(no LHCb D), the fitting range is $0.25\leq x_2 \leq 0.35$.  And we further expanded the range by 50\% for each, serving as a source of systematic uncertainty in our analysis. We will combine these results with the SRC scaling factor $a_2(A)$ measured in lepton inelastic scattering (collected in Tables \ref{slopeandSRC_EPPS21_DY}-\ref{slopeandSRC_nCTEQ15HQ_DY} in App.\,\ref{appendixB}) \cite{CLAS:2019vsb}, which approximately equals the relative abundance of SRC pairs in a nucleus compared to deuteron. Fig.\,\ref{linearDY_figure} shows the slopes versus the SRC scaling factors. The black line corresponds to a fit of numerical results,
\begin{subequations}\label{LinearFitDY}
\begin{eqnarray}
  &&\!\!\!\!\!\!\!\!\!\!\!\!\textrm{\textbf{EPPS21:}} \nn\\
  &&\!\!\!\!\!\!\!\!\!\!\!\!-d R_A^{\textrm{DY}}/dx_2 \!=\! (0.087 \pm 0.003) \Big( a_2(A) - a_2(D) \Big) \,, \\
  &&\!\!\!\!\!\!\!\!\!\!\!\!\textrm{\textbf{nNNPDF3.0(no LHCb D):}} \nn\\
  &&\!\!\!\!\!\!\!\!\!\!\!\!-d R_A^{\textrm{DY}}/dx_2 \!=\! (0.049 \pm 0.012) \Big( a_2(A) - a_2(D) \Big) \,,    \\
  &&\!\!\!\!\!\!\!\!\!\!\!\!\textrm{\textbf{TUJU21:}} \nn\\
  &&\!\!\!\!\!\!\!\!\!\!\!\!-d R_A^{\textrm{DY}}/dx_2 \!=\! (0.067 \pm 0.009) \Big( a_2(A) - a_2(D) \Big) \,, \\
  &&\!\!\!\!\!\!\!\!\!\!\!\!\textrm{\textbf{nCTEQ15HQ:}} \nn\\
  &&\!\!\!\!\!\!\!\!\!\!\!\!-d R_A^{\textrm{DY}}/dx_2 \!=\! (0.081 \pm 0.005) \Big(a_2(A) - a_2(D) \Big) \,.
\end{eqnarray}
\end{subequations}
Here we constrain the fit by forcing the slope to go through zero for the deuteron ($a_2(D)=1$) \cite{Arrington:2012ax}. It can be observed in Fig.\,\ref{linearDY_figure} that the result of EPPS21 exhibits the strongest linear relationship, followed by nCTEQ15HQ and TUJU21, the least favorable is nNNPDF3.0(no LHCb D). In our opinion, the results that fail to reproduce this linear relationship warrant careful re-examination, particularly since the quark-related linearity has been well-established in DIS experiments. Given this prior knowledge, its manifestation in proton-induced Drell-Yan process is also expected. It is worth stressing that the linear relationships exhibited by EPPS21 and nCTEQ15HQ in Fig.\,\ref{linearDY_figure} are remarkably similar to the experimentally established relationship from DIS data \cite{CLAS:2019vsb}, both with and without ISO corrections, as detailed in Fig.\,\ref{combine} in App.\,\ref{appendixC}. This observation is not intended to invalidate the results from the other groups. Rather, it underscores a pronounced tension in the current literature: the discrepancies between the predictions of different groups as seen in Fig.\,\ref{linearDY_figure} are substantial and merit further investigation. The future experimental data over a larger region of $x_2$ ($0.4\leq x_2 \leq 0.7$) will help to clarify this issue.

\section{Summary}
\label{summary}
The EMC effect, which refers to the modification of quark distributions in bound nucleons compared to free ones, has been extensively studied over the past forty years. However, a comprehensive understanding of it is still lacking. How does the nuclear environment plays an important role even at energy scales much higher than those involved in typical nuclear ground-state processes? Although the data do not yet allow for a clear preference between various explanations, the SRC-driven EMC effect offers a natural way out for us. It is the local nuclear structure (SRCs) rather than the global nuclear environment that affects the distributions of partons, and this change is ``averaged'' in the per-nucleon cross section ratio in the DIS experiments.

This study extends the previous work presented in Ref.\,\cite{Wang:2024cpx}. First, focusing on gluons, we utilize four different nPDF parameterizations from EPPS21, nNNPDF3.0(no LHCb D), nCTEQ15HQ, as well as TUJU21, to test the linear relation between the slope of reduced cross section ratio in DIS and the cross section of sub-threshold photoproduction on $J/\psi$. Additionally, we also investigate the linear relationship of quarks in the proton-induced Drell-Yan process with these four parameterizations. Our findings can help deepen the understanding of EMC effect and the universal effects of SRCs on different kinds of partons. Furthermore, the dependence of the results on the nPDF parameterization, as shown in this paper, provide useful reference for more detailed future global fits of nPDFs.

\section{Acknowledgements}
This work is supported in part by National Natural Science Foundation of China under Grant No.  12125503, 12335003, 12105247, 12305106 and 12475098.

\begin{widetext}
\appendix
\section{Collection of $-dR_A^{c \bar{c}}/dx$ and $(\sigma_{A}^{sub}\!/A)$.}
\label{appendixA}
Here we collect the slopes of the ratio factor $-dR_A^{c \bar{c}}/dx$ and the per-nucleon cross sections for sub-threshold $J/\psi$ photoproduction  $(\sigma_{A}^{sub}\!/A)$, to test the linear relation of gluons. The values are calculated by using the four different global analyses of nPDFs.

\begin{table}[H]
  \centering
  \renewcommand{\arraystretch}{1.5}
  \caption{The values of the slopes $-dR_A^{c \bar{c}}/dx$ and the per-nucleon cross sections $(\sigma_{A}^{sub}\!/A)$. These values are calculated by using the global analysis from EPPS21.}\label{slopeandsub_EPPS21}
	\begin{tabular}{c|c c}
	  \hline\hline
	  ~~~Nucleus~~~ & ~~~$ -dR_A^{c \bar{c}}/dx $~~~& ~~~$\sigma_{A}^{sub}\!/A$~~~ \\
	  $^{3}$He   &  $0.135\pm0.010$   & $3.31\pm0.06$  \\
	  $^{4}$He   &  $0.495\pm0.035$     & $11.35\pm0.14$  \\
	  $^{9}$Be   &  $0.587\pm0.040$      &  $13.49\pm0.07$  \\
	  $^{12}$C   &  $0.624\pm0.042$   &  $14.37\pm0.54$  \\
	  $^{27}$Al  &  $0.738\pm0.049$     &  $17.23\pm0.23$ \\
	  $^{56}$Fe  &  $0.857\pm0.056$     &  $20.09\pm0.42$ \\
	  $^{197}$Au &  $1.106\pm0.073$     &  $26.13\pm0.82$  \\
	  \hline\hline
     \end{tabular}
\end{table}	

\begin{table}[H]
  \centering
  \renewcommand{\arraystretch}{1.5}
  \caption{Same as Table \ref{slopeandsub_EPPS21}, except using the global analysis from nNNPDF3.0(no LHCb D).}\label{slopeandsub_nNNPDF}
	\begin{tabular}{c|c c}
	  \hline\hline
	  ~~~Nucleus~~~ & ~~~$ -dR_A^{c \bar{c}}/dx $~~~ & ~~~$\sigma_{A}^{sub}\!/A$~~~ \\
	  $^{4}$He   &  $0.155\pm0.011$   &    $4.17\pm 0.14$  \\
	  $^{9}$Be   &  $0.415\pm0.019$   &    $10.78\pm 0.09$  \\
	  $^{12}$C   &  $0.558\pm0.029$   &    $14.37\pm0.54$     \\
	  $^{27}$Al  &  $1.079\pm0.097$   &    $30.61\pm0.68$ \\
	  $^{56}$Fe  &  $2.279\pm0.073$   &    $53.46\pm0.80$ \\
	  $^{197}$Au &  $3.618\pm0.041$   &    $80.04\pm3.10$  \\
	  \hline\hline
	\end{tabular}
\end{table}	

\begin{table}[H]
  \centering
  \renewcommand{\arraystretch}{1.5}
  \caption{Same as Table \ref{slopeandsub_EPPS21}, except using the global analysis from TUJU21.}\label{slopeandsub_TUJU21}
	\begin{tabular}{c|c c}
	  \hline\hline
	  ~~~Nucleus~~~ & ~~~$ -dR_A^{c \bar{c}}/dx $~~~ & ~~~$\sigma_{A}^{sub}\!/A$~~~ \\
	  $^{4}$He   &  $0.730\pm0.054$   &    $8.77\pm0.12$  \\
	  $^{12}$C   &  $1.070\pm0.109$   &    $14.37\pm0.54$  \\
	  $^{27}$Al  &  $1.240\pm0.135$   &    $17.53\pm0.14$ \\
	  $^{56}$Fe  &  $1.330\pm0.169$   &    $19.83\pm0.27$ \\
	  $^{131}$Xe  &  $1.408\pm0.190$   &    $21.84\pm0.42$ \\
	  $^{197}$Au &  $1.432\pm0.200$  &   $22.56\pm0.48$  \\
	  \hline\hline
	\end{tabular}
\end{table}	

\begin{table}[H]
  \centering
  \renewcommand{\arraystretch}{1.5}
  \caption{Same as Table \ref{slopeandsub_EPPS21}, except using the global analysis from nCTEQ15HQ.}\label{slopeandsub_nCTEQ15HQ}
	\begin{tabular}{c|c c}
	  \hline\hline
	  ~~~Nucleus~~~ & ~~~$-dR_A^{c \bar{c}}/dx $~~~ & ~~~$\sigma_{A}^{sub}\!/A$~~~ \\
	  $^{4}$He   &  $-0.207\pm0.003$   &   $-9.48\pm0.04$  \\
	  $^{9}$Be   &  $-0.273\pm0.006$   &   $-13.36\pm0.02$  \\
	  $^{12}$C   &  $-0.282\pm0.006$    &   $-14.37\pm0.54$  \\
	  $^{27}$Al  &  $-0.243\pm0.008$   &   $-14.80\pm0.10$ \\
	  $^{56}$Fe  &  $-0.074\pm0.009$   &   $-10.20\pm0.26$ \\
	  $^{197}$Au &  $0.645\pm0.020$    &   $19.69\pm0.50$  \\
	  \hline\hline
	\end{tabular}
\end{table}

\section{Collection of $-dR_A^{\textrm{DY}}/dx_2$ and $a_2(A)$.}
\label{appendixB}
Here we collect the slopes of the ratio factor $-dR_A^{\textrm{DY}}/dx_2$ in proton-induced Drell-Yan process and the SRC scaling factors $a_2(A)$, to test the linear relation of quarks. The values are calculated by using the four different global analyses of nPDFs.

\begin{table}[H]
	\centering
	\renewcommand{\arraystretch}{1.5}
	\caption{The slopes of the ratio $-dR_A^{\textrm{DY}}/dx_2$ and the SRC scaling factor $a_2(A)$. These values are calculated by using the global analysis from EPPS21.}\label{slopeandSRC_EPPS21_DY}
	\begin{tabular}{c|c c}
		\hline\hline
		~~~Nucleus~~~ & ~~~ $-dR_A^{\textrm{DY}}/dx_{2}$~~~ & ~~~$a_2(A)$~~~ \\
		$^{3}$He   &  $0.12\pm9\times 10^{-5}$    &   $2.13\pm0.04$  \\
		$^{4}$He   &  $0.218\pm0.007$    &   $3.60\pm0.01$  \\
		$^{9}$Be   &  $0.233\pm0.008$    &   $3.91\pm0.12$  \\
		$^{12}$C   &  $0.271\pm0.009$    &   $4.49\pm0.17$  \\
		$^{27}$Al  &  $0.310\pm0.010$    &   $4.83\pm0.18$ \\
		$^{56}$Fe  &  $0.349\pm0.011$    &   $4.80\pm0.22$ \\
		$^{197}$Au &  $0.416\pm0.014$    &   $5.16\pm0.22$  \\
		\hline\hline
	\end{tabular}
\end{table}	

\begin{table}[H]
  \centering
  \renewcommand{\arraystretch}{1.5}
  \caption{Same as Table \ref{slopeandSRC_EPPS21_DY}, except using the global analysis from nNNPDF3.0(no LHCb D). }\label{slopeandSRC_nNNPDF_DY}
	\begin{tabular}{c|c c}
	  \hline\hline
	  ~~~Nucleus~~~ & ~~~ $-dR_A^{\textrm{DY}}/dx_{2}$~~~ & ~~~$a_2(A)$~~~ \\
	  $^{4}$He   &  $0.018\pm0.022$   &$3.60\pm0.01$  \\
	  $^{9}$Be   &  $0.066\pm0.005$   &$3.91\pm0.12$  \\
	  $^{12}$C   &  $0.085\pm0.003$   &$4.49\pm0.17$  \\
	  $^{27}$Al  &  $0.178\pm0.002$   &$4.83\pm0.18$ \\
	  $^{56}$Fe  &  $0.377\pm0.01$
      &$4.80\pm0.22$ \\
	  $^{197}$Au &  $0.224\pm0.027$   &$5.16\pm0.22$  \\
	  \hline\hline
	\end{tabular}
\end{table}	

\begin{table}[H]
	\centering
	\renewcommand{\arraystretch}{1.5}
	\caption{Same as Table \ref{slopeandSRC_EPPS21_DY}, except using the global analysis from TUJU21.}\label{slopeandSRC_TUJU21_DY}
	\begin{tabular}{c|c c}
		\hline\hline
		~~~Nucleus~~~ & ~~~ $-dR_A^{\textrm{DY}}/dx_{2}$~~~ & ~~~$a_2(A)$~~~ \\
		$^{4}$He   &  $0.079\pm0.003$      &   $3.60\pm0.01$  \\
		$^{12}$C   &  $0.172\pm0.006$    &   $4.49\pm0.17$  \\
		$^{27}$Al  &  $0.234\pm0.005$      &   $4.83\pm0.18$ \\
		$^{56}$Fe  &  $0.286\pm0.004$    &   $4.80\pm0.22$ \\
		$^{197}$Au &  $0.385\pm5.1\times10^{-4}$    &   $5.16\pm0.22$  \\
		\hline\hline
	\end{tabular}
\end{table}	

\begin{table}[H]
	\centering
	\renewcommand{\arraystretch}{1.5}
	\caption{Same as Table \ref{slopeandSRC_EPPS21_DY}, except using the global analysis from nCTEQ15HQ. }\label{slopeandSRC_nCTEQ15HQ_DY}
	\begin{tabular}{c|c c}
		\hline\hline
		~~~Nucleus~~~ & ~~~ $-dR_A^{\textrm{DY}}/dx_{2}$~~~ & ~~~$a_2(A)$~~~ \\
		$^{4}$He   &  $0.145\pm2.3\times10^{-4}$   &   $3.60\pm0.01$  \\
		$^{9}$Be   &  $0.188\pm0.004$   &   $3.91\pm0.12$  \\
		$^{12}$C   &  $0.276\pm0.004$   &   $4.49\pm0.17$  \\
		$^{27}$Al  &  $0.330\pm0.008$   &   $4.83\pm0.18$ \\
		$^{56}$Fe  &  $0.361\pm0.010$   &   $4.80\pm0.22$ \\
		$^{197}$Au &  $0.345\pm0.014$      &   $5.16\pm0.22$  \\
		\hline\hline
	\end{tabular}
\end{table}

\section{Comparison of linear relation with and without ISO corrections.}
\label{appendixC}

To illustrate the effect of ISO corrections, Fig.\,\ref{combine} shows the linear fitting results between the slopes of $R_A^{\textrm{DY}}(x_2, Q^2)$ and the SRC scaling factors $a_2(A)$, both with and without ISO corrections. As observed, the $-d R_A^{\textrm{DY}}/dx_2$ values show an evident reduction after applying the correction. This effect is more pronounced for nuclei with larger neutron-proton number differences.

\begin{figure*}[b]
	\centering
    \vspace{-20pt}
	\subfigure{
		\includegraphics[width=0.75\linewidth]{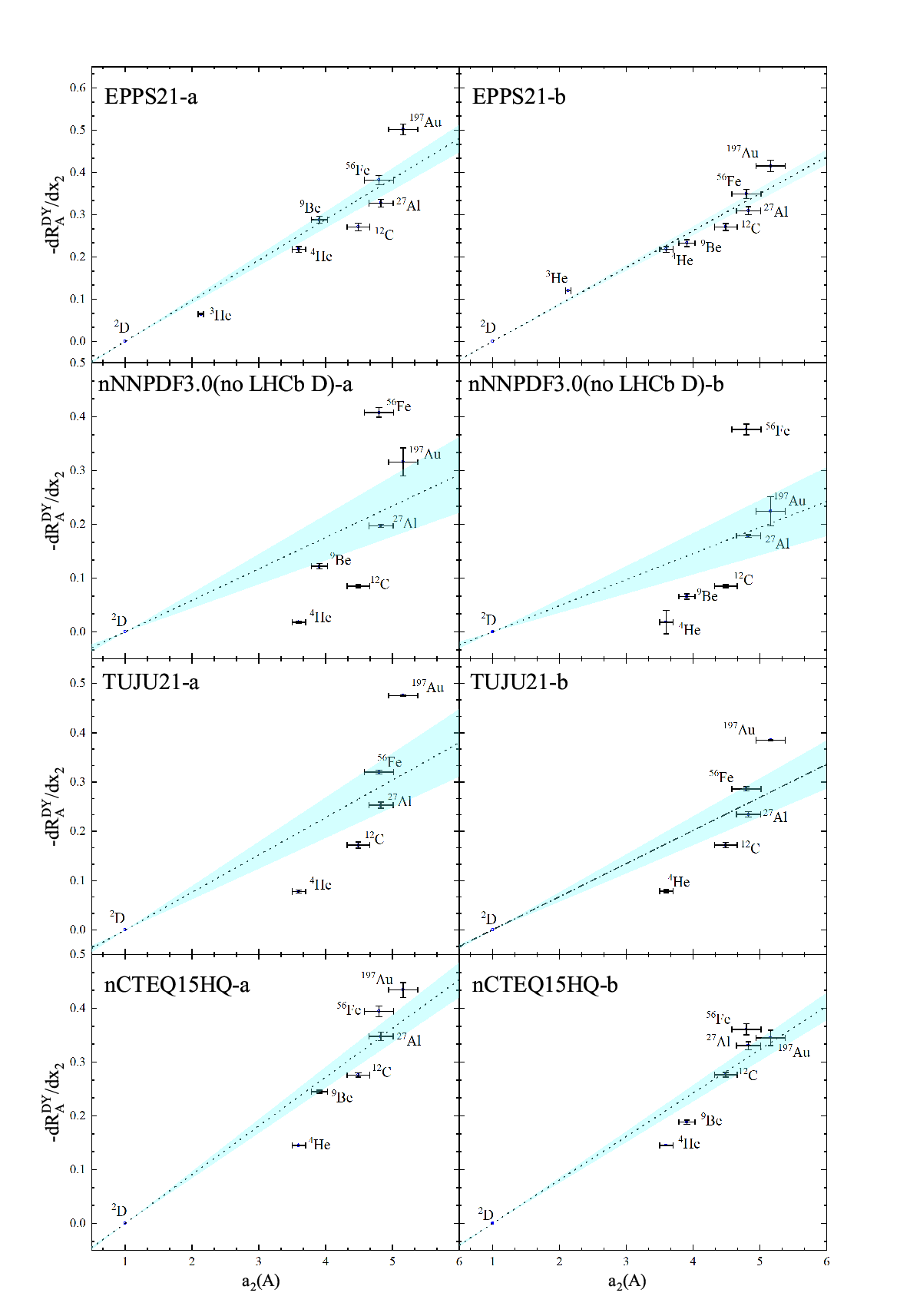}}
	\caption{A comparison of the linear relation for the four different groups with and without the ISO corrections. The left panel shows the results without the ISO corrections (labeled ``a"), while the right panel presents those with the ISO corrections applied (labeled ``b").}
	\label{combine}
\end{figure*}

\end{widetext}


\begin{thebibliography}{}
		
\bibitem{EuropeanMuon:1983wih}
J.~J.~Aubert \textit{et al.} [European Muon],
Phys. Lett. B \textbf{123} (1983), 275-278
doi:10.1016/0370-2693(83)90437-9
		
\bibitem{EuropeanMuon:1988lbf}
J.~Ashman \textit{et al.} [European Muon],
Phys. Lett. B \textbf{202} (1988), 603-610
doi:10.1016/0370-2693(88)91872-2
		
\bibitem{Gomez:1993ri}
J.~Gomez, R.~G.~Arnold, P.~E.~Bosted, C.~C.~Chang, A.~T.~Katramatou, G.~G.~Petratos, A.~A.~Rahbar, S.~E.~Rock, A.~F.~Sill and Z.~M.~Szalata, \textit{et al.}
Phys. Rev. D \textbf{49} (1994), 4348-4372
doi:10.1103/PhysRevD.49.4348
		
\bibitem{EuropeanMuon:1988tpw}
M.~Arneodo \textit{et al.} [European Muon],
Phys. Lett. B \textbf{211} (1988), 493-499
doi:10.1016/0370-2693(88)91900-4
		
\bibitem{NewMuonNMC:1990xyw}
D.~Allasia \textit{et al.} [New Muon (NMC)],
Phys. Lett. B \textbf{249} (1990), 366-372
doi:10.1016/0370-2693(90)91270-L
		
\bibitem{Seely:2009gt}
J.~Seely, A.~Daniel, D.~Gaskell, J.~Arrington, N.~Fomin, P.~Solvignon, R.~Asaturyan, F.~Benmokhtar, W.~Boeglin and B.~Boillat, \textit{et al.}
Phys. Rev. Lett. \textbf{103} (2009), 202301
doi:10.1103/PhysRevLett.103.202301
[arXiv:0904.4448 [nucl-ex]].

\bibitem{CLAS:2019vsb}
B.~Schmookler \textit{et al.} [CLAS],
Nature \textbf{566}, no.7744, 354-358 (2019)
doi:10.1038/s41586-019-0925-9
[arXiv:2004.12065 [nucl-ex]].


		
\bibitem{Wang:2022kwg}
R.~Wang, N.~N.~Ma and T.~F.~Wang,
Chin. Phys. C \textbf{47} (2023) no.4, 044103
doi:10.1088/1674-1137/acb7d0
[arXiv:2207.10980 [nucl-th]].

\bibitem{Bertsch:1993vx}
G.~F.~Bertsch, L.~Frankfurt and M.~Strikman,
Science \textbf{259}, 773-774 (1993)
doi:10.1126/science.259.5096.773

\bibitem{Miller:2013hla}
G.~A.~Miller,
Phys. Rev. C \textbf{89}, no.4, 045203 (2014)
doi:10.1103/PhysRevC.89.045203
[arXiv:1311.4561 [nucl-th]].

\bibitem{Frank:1995pv}
M.~R.~Frank, B.~K.~Jennings and G.~A.~Miller,
Phys. Rev. C \textbf{54}, 920-935 (1996)
doi:10.1103/PhysRevC.54.920
[arXiv:nucl-th/9509030 [nucl-th]].

\bibitem{Wang:2016mzo}
R.~Wang, X.~Chen and Q.~Fu,
Nucl. Phys. B \textbf{920}, 1-19 (2017)
doi:10.1016/j.nuclphysb.2017.04.008
[arXiv:1611.03670 [hep-ph]].

\bibitem{Zhang:2009vj}
Y.~Zhang, L.~Shao and B.~Q.~Ma,
Nucl. Phys. A \textbf{828}, 390-400 (2009)
doi:10.1016/j.nuclphysa.2009.07.006
[arXiv:0909.0454 [nucl-th]].

\bibitem{Weinstein:2010rt}
L.~B.~Weinstein, E.~Piasetzky, D.~W.~Higinbotham, J.~Gomez, O.~Hen and R.~Shneor,
Phys. Rev. Lett. \textbf{106}, 052301 (2011)
doi:10.1103/PhysRevLett.106.052301
[arXiv:1009.5666 [hep-ph]].

\bibitem{Hen:2014nza}
O.~Hen, M.~Sargsian, L.~B.~Weinstein, E.~Piasetzky, H.~Hakobyan, D.~W.~Higinbotham, M.~Braverman, W.~K.~Brooks, S.~Gilad and K.~P.~Adhikari, \textit{et al.}
Science \textbf{346} (2014), 614-617
doi:10.1126/science.1256785
[arXiv:1412.0138 [nucl-ex]].

\bibitem{CLAS:2005ola}
K.~S.~Egiyan \textit{et al.} [CLAS],
Phys. Rev. Lett. \textbf{96} (2006), 082501
doi:10.1103/PhysRevLett.96.082501
[arXiv:nucl-ex/0508026 [nucl-ex]].

\bibitem{Wang:2024ikx}
W.~Wang, J.~Xu, X.~H.~Yang, Y.~T.~Zhang and S.~Zhao,
[arXiv:2409.14367 [hep-ph]].

\bibitem{Hen:2012fm}
O.~Hen, E.~Piasetzky and L.~B.~Weinstein,
Phys. Rev. C \textbf{85} (2012), 047301
doi:10.1103/PhysRevC.85.047301
[arXiv:1202.3452 [nucl-ex]].
	
		
\bibitem{CLAS:2018yvt}
M.~Duer \textit{et al.} [CLAS],
Nature \textbf{560} (2018) no.7720, 617-621
doi:10.1038/s41586-018-0400-z

\bibitem{Wang:2024cpx}
W.~Wang, J.~Xu, X.~H.~Yang and S.~Zhao,
Eur. Phys. J. A \textbf{61} (2025) no.5, 112
doi:10.1140/epja/s10050-025-01588-4
[arXiv:2401.16662 [hep-ph]].

\bibitem{nCTEQ:2023cpo}
A.~W.~Denniston \textit{et al.} [nCTEQ],
Phys. Rev. Lett. \textbf{133} (2024) no.15, 152502
doi:10.1103/PhysRevLett.133.152502
[arXiv:2312.16293 [hep-ph]].

\bibitem{Paakkinen:2025pcw}
P.~Paakkinen,
[arXiv:2510.00252 [hep-ph]].

\bibitem{Xu:2019wso}
J.~Xu and F.~Yuan,
Phys. Lett. B \textbf{801}, 135187 (2020)
doi:10.1016/j.physletb.2019.135187
[arXiv:1908.10413 [hep-ph]].

\bibitem{Hatta:2019ocp}
Y.~Hatta, M.~Strikman, J.~Xu and F.~Yuan,
Phys. Lett. B \textbf{803}, 135321 (2020)
doi:10.1016/j.physletb.2020.135321
[arXiv:1911.11706 [hep-ph]].

\bibitem{Chen:2016bde}
J.~W.~Chen, W.~Detmold, J.~E.~Lynn and A.~Schwenk,
Phys. Rev. Lett. \textbf{119}, no.26, 262502 (2017)
doi:10.1103/PhysRevLett.119.262502
[arXiv:1607.03065 [hep-ph]].

\bibitem{Wang:2020uhj}
X.~G.~Wang, A.~W.~Thomas and W.~Melnitchouk,
Phys. Rev. Lett. \textbf{125}, 262002 (2020)
doi:10.1103/PhysRevLett.125.262002
[arXiv:2004.03789 [hep-ph]].

\bibitem{Yang:2023zmr}
X.~H.~Yang, F.~Huang and J.~Xu,
Phys. Rev. D \textbf{108}, no.5, 053005 (2023)
doi:10.1103/PhysRevD.108.053005
[arXiv:2305.11538 [hep-ph]].

\bibitem{Huang:2021cac}
F.~Huang, J.~Xu and X.~H.~Yang,
Phys. Rev. D \textbf{104}, no.3, 033002 (2021)
doi:10.1103/PhysRevD.104.033002
[arXiv:2103.07873 [hep-ph]].

\bibitem{Eskola:2021nhw}
K.~J.~Eskola, P.~Paakkinen, H.~Paukkunen and C.~A.~Salgado,
Eur. Phys. J. C \textbf{82} (2022) no.5, 413
doi:10.1140/epjc/s10052-022-10359-0
[arXiv:2112.12462 [hep-ph]].

\bibitem{AbdulKhalek:2022fyi}
R.~Abdul Khalek, R.~Gauld, T.~Giani, E.~R.~Nocera, T.~R.~Rabemananjara and J.~Rojo,
Eur. Phys. J. C \textbf{82} (2022) no.6, 507
doi:10.1140/epjc/s10052-022-10417-7
[arXiv:2201.12363 [hep-ph]].
				
\bibitem{Duwentaster:2022kpv}
P.~Duwent\"aster, T.~Je\v{z}o, M.~Klasen, K.~Kova\v{r}\'\i{}k, A.~Kusina, K.~F.~Muzakka, F.~I.~Olness, R.~Ruiz, I.~Schienbein and J.~Y.~Yu,
Phys. Rev. D \textbf{105} (2022) no.11, 114043
doi:10.1103/PhysRevD.105.114043
[arXiv:2204.09982 [hep-ph]].
				
\bibitem{Helenius:2021tof}
I.~Helenius, M.~Walt and W.~Vogelsang,
Phys. Rev. D \textbf{105} (2022) no.9, 094031
doi:10.1103/PhysRevD.105.094031
[arXiv:2112.11904 [hep-ph]].

\bibitem{Buckley:2014ana}
A.~Buckley, J.~Ferrando, S.~Lloyd, K.~Nordstr{\"o}m, B.~Page, M.~R{\"u}fenacht, M.~Sch{\"o}nherr and G.~Watt,
Eur. Phys. J. C \textbf{75} (2015), 132
doi:10.1140/epjc/s10052-015-3318-8
[arXiv:1412.7420 [hep-ph]].

\bibitem{NuSea:1999egr}
M.~A.~Vasilev \textit{et al.} [NuSea],
Phys. Rev. Lett. \textbf{83} (1999), 2304-2307
doi:10.1103/PhysRevLett.83.2304
[arXiv:hep-ex/9906010 [hep-ex]].

\bibitem{Reimer:2016dcd}
P.~E.~Reimer [Fermilab SeaQuest],
EPJ Web Conf. \textbf{113}, 05012 (2016)
doi:10.1051/epjconf/201611305012

\bibitem{Reimer:2011zza}
P.~E.~Reimer [Fermilab SeaQuest],
J. Phys. Conf. Ser. \textbf{295}, 012011 (2011)
doi:10.1088/1742-6596/295/1/012011

\bibitem{NA10:1987hho}
P.~Bordalo \textit{et al.} [NA10],
Phys. Lett. B \textbf{193} (1987), 368
doi:10.1016/0370-2693(87)91253-6

\bibitem{NA3:1981yaj}
J.~Badier \textit{et al.} [NA3],
Phys. Lett. B \textbf{104} (1981), 335
doi:10.1016/0370-2693(81)90137-4
						


\bibitem{Aschenauer:2017oxs}
E.~C.~Aschenauer, S.~Fazio, M.~A.~C.~Lamont, H.~Paukkunen and P.~Zurita,
Phys. Rev. D \textbf{96} (2017) no.11, 114005
doi:10.1103/PhysRevD.96.114005
[arXiv:1708.05654 [nucl-ex]].

\bibitem{Laenen:1992zk}
E.~Laenen, S.~Riemersma, J.~Smith and W.~L.~van Neerven,
Nucl. Phys. B \textbf{392}, 162-228 (1993)
doi:10.1016/0550-3213(93)90201-Y

\bibitem{AbdulKhalek:2021gbh}
R.~Abdul Khalek, A.~Accardi, J.~Adam, D.~Adamiak, W.~Akers, M.~Albaladejo, A.~Al-bataineh, M.~G.~Alexeev, F.~Ameli and P.~Antonioli, \textit{et al.}
Nucl. Phys. A \textbf{1026}, 122447 (2022)
doi:10.1016/j.nuclphysa.2022.122447
[arXiv:2103.05419 [physics.ins-det]].


\bibitem{Hou:2019qau}
T.~J.~Hou, K.~Xie, J.~Gao, S.~Dulat, M.~Guzzi, T.~J.~Hobbs, J.~Huston, P.~Nadolsky, J.~Pumplin and C.~Schmidt, \textit{et al.}
[arXiv:1908.11394 [hep-ph]].
		


\bibitem{Drell:1970wh}
S.~D.~Drell and T.~M.~Yan,
Phys. Rev. Lett. \textbf{25}, 316-320 (1970)
[erratum: Phys. Rev. Lett. \textbf{25}, 902 (1970)]
doi:10.1103/PhysRevLett.25.316

\bibitem{Huang:2025kmd}
F.~Huang, S.~M.~Hu, D.~M.~Li and J.~Xu,
Eur. Phys. J. C \textbf{85} (2025) no.10, 1225
doi:10.1140/epjc/s10052-025-14960-x
[arXiv:2501.07059 [hep-ph]].

\bibitem{Chmaj:1983jq}
T.~Chmaj and K.~J.~Heller,
Acta Phys. Polon. B \textbf{15} (1984), 473
TPJU-18/83.

\bibitem{Moreno:1990sf}
G.~Moreno, C.~N.~Brown, W.~E.~Cooper, D.~Finley, Y.~B.~Hsiung, A.~M.~Jonckheere, H.~Jostlein, D.~M.~Kaplan, L.~M.~Lederman and Y.~Hemmi, \textit{et al.}
Phys. Rev. D \textbf{43}, 2815-2836 (1991)
doi:10.1103/PhysRevD.43.2815

\bibitem{Alde:1990im}
D.~M.~Alde, H.~W.~Baer, T.~A.~Carey, G.~T.~Garvey, A.~Klein, C.~Lee, M.~J.~Leitch, J.~W.~Lillberg, P.~L.~McGaughey and C.~S.~Mishra, \textit{et al.}
Phys. Rev. Lett. \textbf{64} (1990), 2479-2482
doi:10.1103/PhysRevLett.64.2479


\bibitem{Heinrich:1989cp}
J.~G.~Heinrich, C.~E.~Adolphsen, J.~P.~Alexander, K.~J.~Anderson, J.~S.~Conway, J.~E.~Pilcher, A.~Possoz, E.~I.~Rosenberg, C.~Biino and J.~F.~Greenhalgh, \textit{et al.}
Phys. Rev. Lett. \textbf{63} (1989), 356-359
doi:10.1103/PhysRevLett.63.356



\bibitem{AE866E789E772}
E866/E789/E772 web resources http://p25ext.lanl.gov/e866/papers/papers.html




\bibitem{Kulagin:2014vsa}
S.~A.~Kulagin and R.~Petti,
Phys. Rev. C \textbf{90}, no.4, 045204 (2014)
doi:10.1103/PhysRevC.90.045204
[arXiv:1405.2529 [hep-ph]].


\bibitem{Kenyon:1982tg}
I.~R.~Kenyon,
Rept. Prog. Phys. \textbf{45} (1982), 1261
doi:10.1088/0034-4885/45/11/002


\bibitem{Curci:1979am}
G.~Curci and M.~Greco,
Phys. Lett. B \textbf{92} (1980), 175-178
doi:10.1016/0370-2693(80)90331-7



\bibitem{Arrington:2012ax}
J.~Arrington, A.~Daniel, D.~Day, N.~Fomin, D.~Gaskell and P.~Solvignon,
Phys. Rev. C \textbf{86} (2012), 065204
doi:10.1103/PhysRevC.86.065204
[arXiv:1206.6343 [nucl-ex]].








\end{thebibliography}
\end{document}